\documentclass[
reprint,
superscriptaddress,
longbibliography,
 amsmath,amssymb,
 aps,
prx,
 showkeys,
]{revtex4-1}

\usepackage{mathtools}
\usepackage{sidecap}
\usepackage{lipsum}
\usepackage{graphicx}
\usepackage{textgreek}
\usepackage{dcolumn}
\usepackage{bm}
\usepackage{color}
\usepackage[table,xcdraw,dvipsnames]{xcolor}

\newcommand{\koffS}{k_\text{off}^\text{S}}
\newcommand{\koffW}{k_\text{off}^\text{W}}
\newcommand{\koffR}{k_\text{off}^\text{R}}
\newcommand{\kon}{k_\text{on}}
\newcommand{\tauD}{\tau_{D}}
\newcommand{\rhoE}{\rho_{_{\text{E}}}}
\newcommand{\rhoS}{\rho_{_{\text{S}}}}
\newcommand{\rhoR}{\rho_{_{\text{R}}}}
\newcommand{\rhoW}{\rho_{_{\text{W}}}}
\newcommand{\rhoES}{\rho_{_{\text{ES}}}}
\newcommand{\rhoEW}{\rho_{_{\text{EW}}}}
\newcommand{\rhoER}{\rho_{_{\text{ER}}}}
\newcommand{\jS}{j_{_{\text{S}}}}
\newcommand{\lambdaS}{\lambda_{_{\text{S}}}}
\newcommand{\muS}{\Delta \mu}
\newcommand{\vS}{v_\text{S}}
\newcommand{\vR}{v_\text{R}}
\newcommand{\vW}{v_\text{W}}
\newcommand{\kp}{k_{\text{p}}}
\definecolor{kabircolor}{rgb}{.85,.372,.008}
\definecolor{vahecolor}{rgb}{0.44, 0.62, .35}

\begin{document}

\title{Proofreading through spatial gradients}

\author{Vahe Galstyan}
\affiliation{Biochemistry and Molecular Biophysics Option, California Institute of Technology, Pasadena CA}
\author{Kabir Husain}
\affiliation{Department of Physics and the James Franck Institute, University of Chicago, Chicago IL}
\author{Fangzhou Xiao}
\affiliation{Division of Biology and Biological Engineering, California Institute of Technology, Pasadena CA}
\author{Arvind Murugan}
\email{amurugan@uchicago.edu}
\affiliation{Department of Physics and the James Franck Institute, University of Chicago, Chicago IL}
\author{Rob Phillips}
\email{phillips@pboc.caltech.edu}
\affiliation{Division of Biology and Biological Engineering, California Institute of Technology, Pasadena CA}
\affiliation{Department of Physics, California Institute of Technology, Pasadena CA}

\begin{abstract}
Key enzymatic processes in biology use the nonequilibrium error correction mechanism called kinetic proofreading to enhance their specificity. Kinetic proofreading typically requires several de\-dicated structural features in the enzyme, such as a nucleotide hydrolysis site and multiple enzyme--substrate conformations that delay product formation. Such requirements limit the applicability and the adaptability of traditional proofreading schemes. Here, we explore an alternative conceptual mechanism of error correction that achieves delays between substrate binding and subsequent product formation by having these events occur at distinct physical locations. The time taken by the enzyme--substrate complex to diffuse from one location to another is leveraged to discard wrong substrates. This mechanism does not require dedicated structural elements on the enzyme, making it easier to overlook in experiments but also making proofreading tunable on the fly. We discuss how tuning the length scales of enzyme or substrate concentration gradients changes the fidelity, speed and energy dissipation, and quantify the performance limitations imposed by realistic diffusion and reaction rates in the cell. Our work broadens the applicability of kinetic proofreading and sets the stage for the study of spatial gradients as a possible route to specificity.
\end{abstract}

\maketitle

\section{Introduction}

The nonequilibrium mechanism called kinetic proofreading \cite{Hopfield1974, NinioBiochimie1975} is used for reducing the error rates of many biochemical processes important for cell function (e.g., DNA replication \cite{Kunkel2004}, transcription \cite{SydowCramer2009}, translation \cite{Rodnina2001,IeongPNAS2016}, signal transduction \cite{SwainBiophysJ2002}, or pathogen recognition \cite{Mckeithan1995,Goldstein2004,CuiPloS2018}). Proofreading mechanisms operate by inducing a delay  between substrate binding and product formation via intermediate states for the enzyme--substrate complex. Such a delay gives the enzyme multiple chances to release the wrong substrate after initial binding, allowing far lower error rates than what one would expect solely from the binding energy difference between right and wrong substrates.

\begin{figure}[!t]
\includegraphics{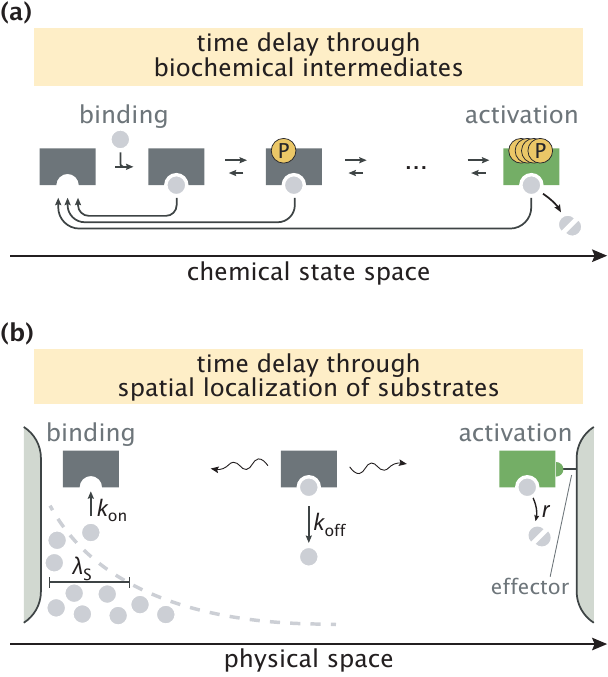}
\caption{
Error correction schemes that operate by delaying product formation.
(a) The traditional proofreading scheme with multiple biochemically distinct intermediates, transitions between which are typically accompanied by energy--consuming reactions.
The T-cell activation mechanism with successive phosphorylation events is used for demonstration \cite{Mckeithan1995, CuiPloS2018}.
(b) The spatial proofreading scheme where the delay between binding and catalysis is created by constraining these events to distinct physical locations. The wavy arrows stand for the diffusive motion of the complex.
Binding events primarily take place on the length scale $\lambdaS$ of substrate localization.
}
\label{fig:intro}
\end{figure}

Traditional proofreading schemes require dedicated molecular features such as an exonuclease pocket in DNA polymerases \cite{Kunkel2004} or multiple phosphorylation sites on T-cell receptors \cite{Mckeithan1995, Goldstein2004}; such features create intermediate states that delay product formation (Fig.~\ref{fig:intro}a) and thus allow proofreading.
Additionally, since proofreading is an active nonequilibrium process often involving near--irreversible reactions,
the enzyme typically needs to have an ATP or GTP hydrolysis site to enable 
the use of energy supplies of the cell \cite{YamaneHopfield1977, Rodnina2001}.
Due to such stringent structural requirements, the number of confirmed proofreading enzymes is relatively small.
Furthermore, generic enzymes without such dedicated features are assumed to not have active error correction available to them.

In this work, we propose an alternative scheme where the delay between initial substrate binding and product formation steps is achieved by separating these events in space. If substrates are spatially localized and product formation is favorable only in a region of low substrate concentration where an activating effector is present, then the time taken by the enzyme--substrate complex to travel from one location to the other can be used to discard the wrong substrates (Fig.~\ref{fig:intro}b). When this delay is longer than substrate unbinding time scales, very low error rates of product formation can be achieved, allowing this spatial proofreading scheme to outperform biochemical mechanisms with a finite number of proofreading steps.

The nonequilibrium mechanism here does not require any direct energy consumption by the enzyme or substrate itself (e.g., through ATP hydrolysis). Instead, the mechanism relies on energy investment to actively maintain spatial concentration gradients of substrates (or alternatively, the enzyme). Such gradients of different proteins in the cell have been measured in several contexts (e.g., near the plasma membrane, the Golgi apparatus, the endoplasmic reticulum (ER), kinetochores, microtubules \cite{Bivona2003-nj, Caudron2005-by, Kholodenko2006-hy}) and several gradient--forming mechanisms have been discussed in the literature \cite{Wu2018-dt, Kholodenko2006-hy, Kholodenko2003--tr}. In this way, energy consumption for proofreading can be outsourced from the enzyme and substrate to the gradient maintaining mechanism.

The scheme proposed here does not rely on any proofreading--specific structural features in the enzyme; indeed, any `equilibrium' enzyme with a localized effector can proofread using our scheme if appropriate concentration gradients of the substrates or enzymes can be set up. As a result, spatial proofreading is easy to overlook in experiments and suggests another explanation for why reconstitution of reactions \emph{in vitro} can be of lower fidelity than \emph{in vivo}.

Further, the lack of reliance on structure makes spatial proofreading more adaptable. We study how tuning the length scale of concentration gradients can trade off error rate against speed and energy consumption on the fly. In contrast, traditional proofreading schemes rely on nucleotide chemical potentials, e.g., the out of equilibrium [ATP]/[ADP] ratio in the cell, and cannot modulate their operation without broader physiological disruptions. We conclude by quantifying the limitations of our proposed scheme by accounting for realistic reaction rates and spatial gradients known to be maintained in the cell. Our work motivates a detailed investigation of spatial structures and compartmentalization in living cells as possible delay mechanisms for proofreading enzymatic reactions.

\section{Results}

\subsection{Slow Transport of Enzymatic Complex Enables Proofreading}

Our proposed scheme is based on spatially separating substrate binding and product formation events for the enzyme (Fig. 1b). Such a setting arises naturally if substrates are spatially localized by having concentration gradients in a cellular compartment. Similarly, an effector needed for product formation (e.g., through allosteric activation) may have a spatial concentration gradient localized elsewhere in that compartment. 
To keep our model simple, we assume that the right (R) and wrong (W) substrates have identical concentration gradients of length scale $\lambdaS$ but that the effector is entirely localized to one end of the compartment, e.g., via membrane tethering.

We model our system using coupled reaction--diffusion equations for the substrate--bound (``ES'' with $\text{S} = \text{R},\text{W}$) and free (``E'') enzyme densities, namely,
\begin{align}
    \label{eqn:ES_pde}
    \frac{\partial \rhoER}{\partial t} &= 
    D \frac{\partial^2 \rhoER}{\partial x^2} 
    - \koffR \rhoER
    + \kon \rhoR \rhoE, \\
    \frac{\partial \rhoEW}{\partial t} &= 
    D \frac{\partial^2 \rhoEW}{\partial x^2} 
    - \koffW \rhoEW
    + \kon \rhoW \rhoE, \\
    \label{eqn:E_pde}
    \frac{\partial \rhoE}{\partial t} &= 
    D \frac{\partial^2 \rhoE}{\partial x^2}
    + \sum_{\text{S}=\text{R},\text{W}} \koffS \rhoES 
    - \sum_{\text{S}=\text{R},\text{W}} \kon \rhoS \rhoE.
\end{align}
Here, $D$ is the enzyme diffusion constant, $\kon$ and $\koffS$ (with $\koffW > \koffR$) are the substrate binding and unbinding rates, respectively, and $\rhoS(x) \sim e^{-x/\lambdaS}$
is the spatially localized substrate concentration profile which we take to be exponentially decaying,
which is often the case for profiles created by cellular gradient formation mechanisms \cite{Driever1988--ti, Brown1999-ew}. We limit our discussion to this one-dimensional setting of the system, though our treatment can be generalized to two and three dimensions in a straightforward way.

The above model does not explicitly account for several effects relevant to living cells, such as depletion of substrates or distinct diffusion rates for the free and substrate--bound enzymes. More importantly, it does not account for the mechanism of substrate gradient formation.
We analyze a biochemically detailed model with this latter feature and experimentally constrained parameters later in the paper.
Here, we proceed with the minimal model above for explanatory purposes.
To identify the key determinants of the model's performance, we assume throughout our analysis  that the amount of substrates is sufficiently low that the enzymes are mostly free with a roughly uniform profile (i.e., $\rhoE \approx \text{constant}$). This assumption makes Eqs.~(\ref{eqn:ES_pde})-(\ref{eqn:E_pde}) linear and allows us to solve them analytically at steady state. We demonstrate in Appendix C that proofreading is, in fact, most effective under this assumption and discuss the consequences of having high substrate amounts on the performance of the scheme.

In our simplified picture, enzyme activation and catalysis take place upon reaching the right boundary at a rate $r$ that is identical for both substrates. Therefore, the density of substrate--bound enzymes at the right boundary can be taken as a proxy for the rate of product formation $\vS$, since
\begin{align}
\vS =  r \rhoES(L),
\end{align} 
where $L$ is the size of the compartment.

To demonstrate the proofreading capacity of the model, we 
first analyze the limiting case where substrates are highly localized to the left end of the compartment ($\lambdaS \ll L$). In this limit, 
the fidelity $\eta$, defined as the number of right products formed per single wrong product, becomes
\begin{align}
\label{eq:errorExpression}
\eta = \frac{\vR}{\vW} = \sqrt{\eta_\text{eq}} \, \frac{\sinh \left( \sqrt{ \tauD k^{\text{W}}_{\text{off}} } \right)}{ \sinh \left( \sqrt{ \tauD k^{\text{R}}_{\text{off}} } \right) },
\end{align}
where $\eta_\text{eq} = \koffW/\koffR$ is the equilibrium fidelity,
and $\tauD = L^2/D$ is the characteristic time scale of diffusion across the compartment (see Appendix A for the derivation).

\begin{figure}
    \includegraphics{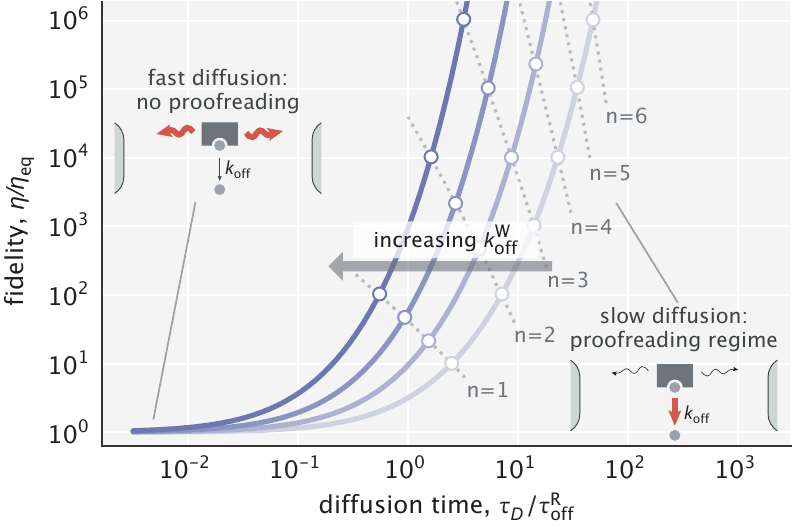}
    \caption{
    Dependence of fidelity on the diffusion time scale in the limit of very high substrate localization. Individual curves were made for different choices of $\koffW$ (varied in the $[10 - 100]\, \koffR$ range).
    $\tau^{\text{R}}_{\text{off}} = 1/\koffR$ is the unbinding time scale of right substrates, kept fixed in the study.
    Fidelity values corresponding to integer degrees of proofreading in a traditional sense ($\eta/\eta_{\text{eq}} = \eta_{\text{eq}}^n$, $n = 1,2,3,...$) are marked as circles. Dominant processes in the two limiting regimes are highlighted in red in the schematics shown as insets.
    }
    \label{fig:intuition}
\end{figure}

Eq.~\ref{eq:errorExpression} is plotted in Fig.~\ref{fig:intuition} for a family of different parameter values.
As can be seen, when diffusion is fast (small $\tauD$), fidelity converges to its equilibrium value and proofreading is lost ($\eta \approx \sqrt{\eta_{\text{eq}}} \times \scalebox{0.8}{$\sqrt{\tauD \koffW/\tauD \koffR}$} = \eta_{\text{eq}}$).
Conversely, when diffusion is slow (large $\tauD$), the enzyme undergoes 
multiple rounds of binding and unbinding before diffusing across the compartment and forming a product -- `futile cycles' that endow the system with proofreading. In this regime, fidelity scales as
\begin{align}
\eta \sim e^{\left(\sqrt{\koffW} - \sqrt{\koffR} \right) \sqrt{\tauD}}.
\end{align}

To get further insights, we introduce an effective number of extra biochemical intermediates ($n$) that a traditional proofreading scheme would need to have in order to yield the same fidelity, i.e., $\eta/\eta_{\text{eq}} = \eta_{\text{eq}}^n$. We calculate this number as (see Appendix A)
\begin{align}
    \label{eqn:effective_n}
    n \approx \frac{\sqrt{\tauD \koffW}}{\ln \eta_{\text{eq}}}.
\end{align}
Notably, since $\tauD \sim L^2$, the result above suggests a linear relationship between the effective number of proofreading realizations and the compartment size ($n \sim L$). In addition, because the right-hand side of Eq.~\ref{eqn:effective_n} is an increasing function of $\koffW$, the proofreading efficiency of the scheme rises with larger differences in substrate off-rates (Fig.~\ref{fig:intuition}) -- a feature that `hard--wired' traditional proofreading schemes lack.

\begin{figure*}[!ht]
\includegraphics[width=0.95\textwidth]{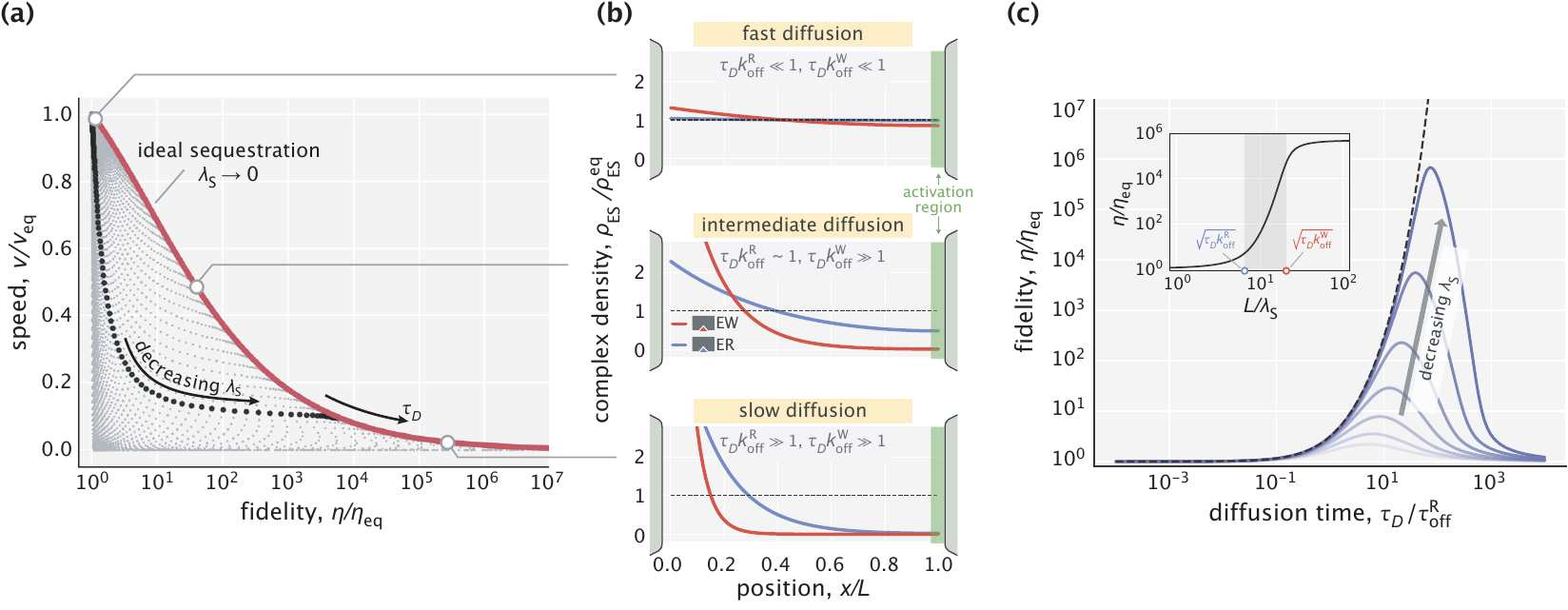}
\caption{
Speed--fidelity trade-off and consequences of having weak substrate gradients.
(a) Speed and fidelity evaluated for sampled values of the diffusion time scale ($\tauD$) and substrate localization length scale ($\lambdaS$). The red line corresponds to the Pareto--optimal front, and is reached in the high substrate localization limit.
(b) Density profiles of wrong (EW) and right (ER) complexes in three qualitatively different performance regimes. The normalization factor $\rhoES^{\text{eq}}$ corresponds to the equilibrium complex densities.
(c) Fidelity as a function of diffusion time scale for different choices of $\lambdaS$. The dashed line corresponds to the ideal sequestration limit ($\lambdaS \rightarrow 0$). Inset: Fidelity as a function of $L/\lambdaS$ for a fixed $\tauD$. Shaded area indicates the range where the bulk of fidelity enhancement takes place.
Equilibrium fidelity $\eta_{\text{eq}} = 10$ was used in generating all the panels.}
\label{fig:trade_off}
\end{figure*}

\subsection*{Navigating the Speed--Fidelity Trade-Off}

As is inherent to all proofreading schemes, the fidelity enhancement described earlier comes at a cost of reduced product formation speed. This reduction, in our case, happens because of increased delays in diffusive transport. Here, we explore the resulting speed--fidelity trade-off and its different regimes by varying two of the model parameters: diffusion time scale $\tauD$ and the substrate localization length scale $\lambdaS$.

Speed and fidelity for different sampled values of $\tauD$ and $\lambdaS$ are depicted in Fig.~\ref{fig:trade_off}a. 
As can be seen, for a fixed $\tauD$, the reduction of $\lambdaS$ can trade off fidelity against speed. This trade-off is intuitive; with tighter substrate localization, the complexes are formed closer to the left boundary. Hence, a smaller fraction of complexes reach the activation region, reducing reaction speed. The Pareto--optimal front of the trade-off over the whole parameter space, shown as a red curve on the plot, is reached in the limit of ideal sequestration.
Varying the diffusion time scale allows one to navigate this optimal trade-off curve and access different performance regimes.

Specifically, if the diffusion time scale is fast compared with the time scales of substrate unbinding (i.e., $\tauD \ll 1/\koffR, 1/\koffW$), then both right and wrong complexes that form near the left boundary arrive at the activation region with high probability, resulting in high speeds, though at the expense of error--prone product formation (Fig.~\ref{fig:trade_off}b, top).
In the opposite limit of slow diffusion, both types of complexes have exponentially low densities at the activation region, but due to the difference in substrate off-rates, production is highly accurate (Fig.~\ref{fig:trade_off}b, bottom).
There also exists an intermediate regime where a significant fraction of right complexes reach the activation region while the vast majority of wrong complexes do not (Fig.~\ref{fig:trade_off}b, middle). As a result, an advantagenous trade-off is achieved where a moderate decrease in the production rate yields high fidelity enhancement -- a feature that was also identified in multi-step traditional proofreading models \cite{MuruganPNAS2012}.

As we saw in Fig.~\ref{fig:trade_off}a, in the case of ideal sequestration, the slowdown of diffusive transport necessarily reduced the production rate and increased the fidelity. The latter part of this statement, however, breaks down when substrate gradients are weak. Indeed, fidelity exhibits a non-monotonic response to tuning $\tauD$ when the substrate gradient length scale $\lambdaS$ is non-zero (Fig.~\ref{fig:trade_off}c). The reason for the eventual decay in fidelity is the fact that with slower diffusion (larger $\tauD$), substrate binding and unbinding events take place more locally and therefore, the right and wrong complex profiles start to resemble the substrate profile itself, which does not discriminate between the two substrate kinds.

Not surprisingly, the error--correcting capacity of the scheme improves with better substrate localization (lower $\lambdaS$). For a fixed $\tauD$, the bulk of this improvement takes place when 
$L/\lambdaS$ is tuned in a range set by the two key dimensionless numbers of the model, namely, $\sqrt{\tauD \koffR}$ and $\sqrt{\tauD \koffW}$ (Fig.~\ref{fig:trade_off}c, inset). 
In Appendix A, we provide an analytical justification for this result.
Taken together, these parametric studies uncover the operational principles of the spatial proofreading scheme and demonstrate how the speed--fidelity trade-off could be dynamically navigated as needed by tuning the key time and length scales of the model.

\subsection*{Energy Dissipation and Limits of Proofreading Performance}

A hallmark signature of proofreading is that it is a nonequilibrium mechanism with an associated free energy cost. In our scheme, the enzyme itself is not directly involved in any energy--consuming reactions, such as hydrolysis. Instead, the free energy cost comes from maintaining the spatial gradient of substrates, which the enzymatic reaction tends to homogenize by releasing bound substrates in regions of low substrate concentration.

While mechanisms of gradient maintenance may differ in their energetic efficiency, there exists a thermodynamically dictated minimum energy that any such mechanism must dissipate per unit time. We calculate this minimum power $P$ as
\begin{align}
\label{eqn:power}
P &= \sum_{\text{S = \{R,W\}}}\int_{0}^L \jS(x) \mu(x) \, \mathrm{d}x.
\end{align}
Here $\jS(x) = \kon \rhoS(x) \rhoE - \koffS \rhoES(x)$ is the net local binding flux of substrate ``S'', and $\mu(x)$ is the local chemical potential (see Appendix B1 for details).
For substrates with an exponentially decaying profile considered here, the chemical potential is given by
\begin{align}
    \mu (x) = \mu(0) + k_{\text{B}}T \ln \frac{\rhoS(x)}{\rhoS(0)} = \mu(0) - k_{\text{B}}T \frac{x}{\lambdaS},
\end{align}
where $k_{\text{B}} T$ is the thermal energy scale.
Notably, the chemical potential difference across the compartment, which serves as an effective driving force for the scheme, is set by the inverse of the nondimensionalized substrate localization length scale, namely,
\begin{align}
    \beta \muS = \frac{L}{\lambdaS},
\end{align}
where $\beta^{-1} = k_{\text{B}}T$.
This driving force is zero for a uniform substrate profile ($\lambdaS \rightarrow \infty$) and increases with tighter localization (lower $\lambdaS$), as intuitively expected.

\begin{figure}
    \centering
    \includegraphics{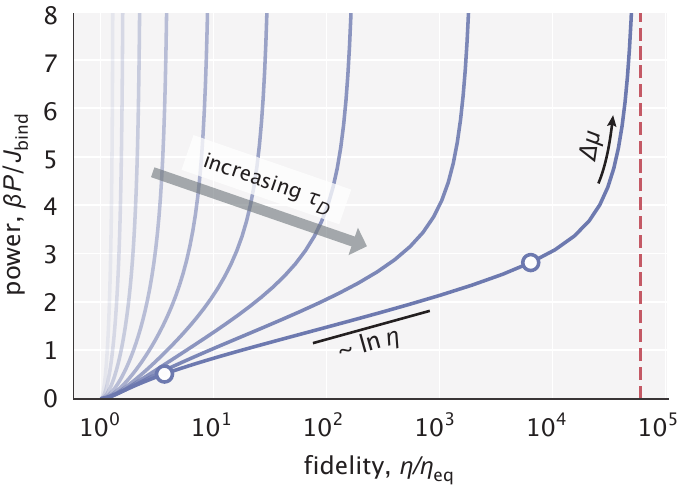}
    \caption{
    Power--fidelity relationship when tuning the effective driving force $\muS$ for different choices of the diffusion time scale $\tauD$. 
    $J_{\text{bind}} = \kon \rhoE \int \rhoS(x) \, \mathrm{d}x$ is the integrated rate of substrate binding (see Appendix B for details).
    The red line indicates the large dissipation limit of fidelity given by Eq.~\ref{eq:errorExpression}. The two circles indicate the $\Delta \mu$ range specified in Eq.~\ref{eqn:dmu_range}.}
    \label{fig:cost}
\end{figure}

We used Eq.~\ref{eqn:power} to study the relationship between dissipation and fidelity enhancement as we tuned $\muS$ for different choices of the diffusion time scale $\tauD$. As can be seen in Fig.~\ref{fig:cost}, power rises with increasing fidelity, diverging when fidelity reaches its asymptotic maximum given by Eq.~\ref{eq:errorExpression} in the large $\muS$ limit. For the bulk of each curve, power scales as the logarithm of fidelity, suggesting that a linear increase in dissipation can yield an exponential reduction in error.
Notably, such a scaling relationship has also been proposed for a general class of biochemical processes involving quality control \cite{HorowitzPRE2017}.
This logarithmic scaling is achieved in our model when the driving force is in a range where most of the fidelity enhancement takes place, namely,
\begin{align}
    \label{eqn:dmu_range}
    \beta \muS \in \left[\sqrt{\tauD \koffR}, \sqrt{\tauD \koffW} \right].
\end{align}
Beyond this range, additional error correction is attained at an increasingly higher cost.

Note that the power computed here does not include the baseline cost of creating the substrate gradient, which, for instance, would depend on the substrate diffusion constant. We only account for the additional cost to be paid due to the operation of the proofreading scheme which works to homogenize this substrate gradient. The baseline cost in our case is analogous to the work that ATP synthase needs to perform to maintain a nonequilibrium [ATP]/[ADP] ratio in the cell, whereas our calculated power is analogous to the rate of ATP hydrolysis by a traditional proofreading enzyme. 
We discuss the comparison between these two classes of dissipation in greater detail in Appendix B3.

Just as the cellular chemical potential of ATP or GTP imposes a thermodynamic upper bound on the fidelity enhancement by any proofreading mechanism \cite{Qian2006}, the effective driving force $\muS$ imposes a similar constraint for the spatial proofreading model. This thermodynamic limit depends only on the available chemical potential and is equal to $e^{\beta \muS}$.
This limit can be approached very closely by our model, which for $\Delta \mu \gtrsim 1$ achieves the exponential enhancement with an additional linear prefactor, namely, $(\eta/\eta_\text{eq})^\text{max} \approx e^{\beta \muS} / \beta \Delta \mu$ (see Appendix B2).
Such scaling behavior was theoretically accessible only to infinite--state traditional proofreading schemes \cite{Qian2006, Ehrenberg1980}.
This offers a view of spatial proofreading as a procession of the enzyme through an infinite series of spatial filters
and suggests that, from the perspective of peak error reduction capacity, our model outperforms the finite--state schemes.

\begin{SCfigure*}
    \centering
    \includegraphics[scale = 1.00]{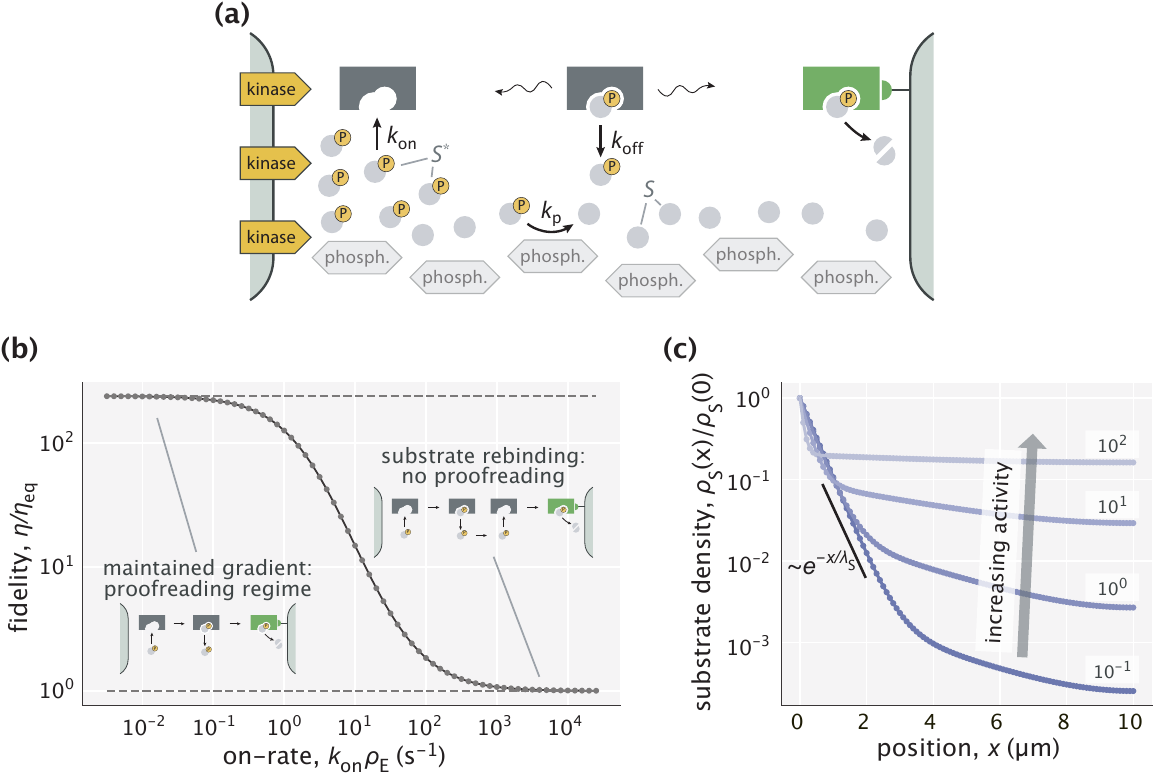}
    \caption{Proofreading based on substrate gradients formed by spatially separated kinases and phosphatases.
    (a) The active form $S^*$ of many proteins exhibits gradients because kinases that phosphorylate $S$ are anchored to a membrane while phosphatases can diffuse in the cytoplasm \cite{Kholodenko2006-hy}.
    An enzyme can exploit the resulting spatial gradient for proofreading.
    (b) At low enzyme activity (i.e., low $\kon \rhoE$), the gradient of $S^*$ is successfully maintained, allowing for proofreading. At high enzyme activity (large $\kon\rhoE$), the
    dephosphorylation with rate
    $\kp = 5\, \text{s}^{-1}$ is no longer sufficient to maintain the gradient and proofreading is lost. (c) Substrate profiles for different choices of enzyme activity. Numbers indicate $\kon\rhoE$ in $\text{s}^{-1}$ units. The black line shows an exponential substrate profile with a length scale $\lambdaS = \sqrt{D/\kp} \sim 0.5\, \text{\textmu m}$.}
    \label{fig:biomodel}
\end{SCfigure*}

\subsection*{Proofreading by Biochemically Plausible Intracellular Gradients}

Our discussion of the minimal model thus far was not aimed at a particular biochemical system and thus did not involve the use of realistic reaction rates and diffusion constants typically seen in living cells. Furthermore, we did not account for the possibility of substrate diffusion, as well as for the homogenization of substrate concentration gradients due to enzymatic reactions, and have thereby abstracted away the gradient maintaining mechanism. The quantitative inspection of such mechanisms is important for understanding the constraints on spatial proofreading in realistic settings.

Here, we investigate proofreading based on a widely applicable mechanism for creating gradients by the spatial separation of two opposing enzymes \cite{Stelling2009-wd, Bivona2003-nj,Brown1999-ew}.
Consider a protein $S$ that is phosphorylated by a membrane--bound kinase and dephosphorylated by a delocalized cytoplasmic phosphatase, as shown in Fig.~\ref{fig:biomodel}a. This setup will naturally create a gradient of the active form of protein ($S^*$), with the gradient length scale controlled by the rate of phosphatase activity $\kp$ ($S^* \xrightarrow{\kp} S$). Such mechanisms are known to create gradients of the active forms of MEK and ERK \cite{Kholodenko2006-hy}, of GTPases such as Ran (with GEF and GAP \cite{Kalab2002-yq} playing the role of kinase and phosphatase, respectively), of cAMP \cite{Kholodenko2006-hy} and of stathmin oncoprotein 18 (Op18) \cite{Bastiaens2006-li,Niethammer2004-at} near the plasma membrane, the Golgi apparatus, the ER, kinetochores and other places.

We test the proofreading power of such gradients, assuming experimentally constrained biophysical parameters for the gradient forming mechanism. Specifically, we consider an enzyme $E$ that acts on active forms of cognate ($R^*$) and non-cognate ($W^*$) substrates which have off-rates $0.1\, \text{s}^{-1}$ and $1\, \text{s}^{-1}$, respectively (hence, $\eta_{\text{eq}} = 10$). These off-rates are consistent with typical values for substrates proofread by cellular signalling systems \cite{CuiPloS2018, Gascoigne2001}.
We assume that both $R^*$ and $W^*$ have identical spatial gradients due to the kinase/phosphatase setup shown in Fig.~\ref{fig:biomodel}a (i.e., $S$ represents both $R$ and $W$). We then consider a dephosphorylation rate constant $\kp = 5\, \text{s}^{-1}$ that falls in the range $0.1 - 100\, \text{s}^{-1}$ reported for different phosphatases \cite{Brown1999-ew, Kholodenko2000-ly,Todd1999-hd}, and a cytosolic diffusion constant $D = 1\, \text{\textmu m}^2/\text{s}$ for all proteins in this model. With this setup, exponential gradients of length scale $\sim0.5\, \text{\textmu m}$ are formed for $R^*$ and $W^*$ (see Appendix D for details).

As expected, proofreading by these gradients is most effective when the enzyme--substrate binding is very slow, in which case the exponential substrate profile is maintained and the system attains the fidelity predicted by our earlier explanatory model (Fig.~\ref{fig:biomodel}b). The system's proofreading capacity is retained if the first--order on-rate is raised up to $\kon \rhoE \sim 10 \, \text{s}^{-1}$,
where around {10-fold} increase in fidelity is still possible.
If the binding rate constant ($\kon$) or the enzyme's expression level ($\rhoE$) is any higher, then enzymatic reactions overwhelm the ability of the kinase/phosphatase system to keep the active forms of substrates sufficiently localized (Fig.~\ref{fig:biomodel}c) and proofreading is lost.
Overall, this model suggests that enzymes can work at reasonable binding rates and still proofread, when accounting for an experimentally characterized gradient maintaining mechanism.

\section*{Discussion}

We have outlined a way for enzymatic reactions to proofread and improve specificity by exploiting spatial concentration gradients of substrates. Like the classic model, our proposed spatial proofreading scheme is based on a time delay; but unlike the classic model, here the delay is due to spatial transport rather than transitions through biochemical intermediates. Consequently, the enzyme is liberated from the stringent structural requirements imposed by traditional proofreading, such as multiple intermediate conformations and hydrolysis sites for energy coupling. Instead, our scheme exploits the free energy supplied by active mechanisms that maintain spatial structures.

The decoupling of the two crucial features of proofreading -- time delay and free energy dissipation -- allows the cell to tune proofreading on the fly. For instance, all proofreading schemes offer fidelity at the expense of reaction speed and energy. For traditional schemes, navigating this trade-off is not always feasible, as it needs to involve structural changes via mutations or modulation of the [ATP]/[ADP] ratio which can cause collateral effects on the rest of the cell. In contrast, the spatial proofreading scheme is more adaptable to the changing conditions and needs of the cell. The scheme can prioritize speed in one context, and fidelity in another, simply by tuning the length scale of intracellular gradients (e.g., through the regulation of the phosphotase or free enzyme concentration in the scheme discussed earlier).

On the other hand, this modular decoupling can complicate the experimental identification 
of proofreading enzymes and the interpretation of their fidelity.
Here, the enzymes need not be endowed with 
the structural and biochemical properties typically sought for in a proofreading enzyme.
At the same time, any attempt to reconstitute enzymatic activity in a well--mixed, \emph{in vitro} assay, will show poor fidelity compared to \emph{in vivo} measurements, even when all necessary molecular players are present \emph{in vitro}.
Therefore, more care is required in studies of cellular information processing mechanisms that hijack a distant source of free energy compared to the case where the relevant energy consumption is local and easier to link causally to function.

While we focused on spatially localized substrates and delocalized enzymes, our framework would apply equally well to other scenarios, e.g., a spatially localized enzyme (or its active form \cite{Kalab2002-yq, Nalbant2004-jb}) and effector with delocalized substrates. Our framework can also be extended to signaling cascades, where slightly different phosphatase activities can result in magnified concentration ratios of two competing signaling molecules at the spatial location of the next cascade step \cite{Roy2009-qh, Bauman2002-za, Kholodenko2006-hy}.

The spatial gradients needed for the operation of our model can be created and maintained through multiple mechanisms in the cell, ranging from the kinase/phosphatase system modeled here, to the passive diffusion of substrates/ligands combined with active degradation (e.g., Bicoid and other developmental morphogens), to active transport processes combined with diffusion. A particularly simple implementation of our scheme is via compartmentalization -- substrates and effectors need to be localized in two spatially separated compartments with the enzyme--substrate complex having to travel from one to another to complete the reaction. 
As specificity is known to be a critical problem in secretory pathways involving the naturally compartmentalized parts of the cell, e.g., the ER, the Golgi apparatus with its distinct cisternae, endosomes and the plasma membrane \cite{Ellgaard2003-wr, Arvan2002-gn}, they are potential candidates for the implementation of spatial proofreading. Experimental investigations of these compartmentalized structures in light of our work will reveal the extent to which spatial transport promotes specificity.

In conclusion, we have analyzed the role played by spatial structures in endowing enzymatic reactions with kinetic proofreading. Simply by spatially segregating substrate binding from catalysis, enzymes can enhance their specificity. This suggests that enzymatic reactions may acquire \textit{de-novo} proofreading capabilities by coupling to pre-existing spatial gradients in the cell.

\begin{acknowledgments}
We thank Anatoly Kolomeisky and Erik Winfree for insightful discussions, and Soichi Hirokawa for providing useful feedback on the manuscript. We also thank Alexander Grosberg who's idea of a compartmentalized `rotary demon' motivated the development of our model. This work was supported by the NIH Grant 1R35 GM118043-01, the John Templeton Foundation Grants 51250 and 60973 (to R.P.), a James S. McDonnell Foundation postdoctoral fellowship (to K.H.), and the Simons Foundation (A.M.).
\end{acknowledgments}



\bibliography{papers.bib}
\end{document}


\title{Proofreading through spatial gradients: \\ Supporting information}    

\author{Vahe Galstyan}
\author{Kabir Husain}
\author{Fangzhou Xiao}
\author{Arvind Murugan}
\author{Rob Phillips}

\maketitle

\onecolumngrid
\tableofcontents

\appendix

\setcounter{equation}{0}
\setcounter{figure}{0}
\setcounter{page}{1}
\renewcommand{\thepage}{S\arabic{page}}
\renewcommand{\thefigure}{S\arabic{figure}}
\renewcommand{\thetable}{S\arabic{table}}
\renewcommand{\theequation}{S\arabic{equation}}

\section*{Appendix A: Analytical calculations of the complex density profile and fidelity}

We begin this section by deriving an analytical expression for the density profile of substrate--bound enzymes ($\rhoES(x)$) in the case where the $\rho(x) \approx \text{constant}$ assumption holds. Based on this result, we then obtain expressions for fidelity in low, high, and intermediate substrate localization regimes. We reserve the studies of speed and fidelity in the general case of a nonuniform free enzyme profile to Appendix C.

\subsection{1. Derivation of the complex density profile $\rhoES(x)$}
The ordinary differential equation (ODE) that defines the steady state profile of substrate--bound enzymes is
\begin{align}
    \label{eqnS:main}
    \underbrace{D \frac{\mathrm{d}^2 \rhoES}{\mathrm{d} x^2}}_{\text{diffusion}} - \underbrace{\koffS \rhoES(x)}_{\text{unbinding}} + \underbrace{\kon  \rhoS(0) e^{-x/\lambdaS} \rhoE(x)}_{\text{binding}} = 0.
\end{align}
Here $\rhoS(0)$ is the substrate density at the leftmost boundary, whose value can be calculated from the condition that the total number of free substrates is $S_{\text{total}}$, namely,
\begin{align}
    S_\text{total} &= \int_{x=0}^L \rhoS(0) e^{-x/\lambdaS} \, \mathrm d x \nonumber\\
    &=  \rhoS(0) \lambdaS \left( 1 - e^{-L/\lambdaS} \right) \Rightarrow  \\
    \label{eqn:rhoS0}
    \rhoS(0) &= \frac{S_\text{total}}{\lambdaS \left( 1 - e^{-L/\lambdaS} \right)}.
\end{align}
In the limit of low substrate amounts where the approximation
$\rhoE(x) \approx \text{constant}$ is valid, Eq.~\ref{eqnS:main}
represents a linear nonhomogeneous ODE.
Hence, its solution can be written as
\begin{align}
    \rhoES(x) = \rhoES^\text{(h)}(x) + \rhoES^\text{(p)}(x),
\end{align}
where $\rhoES^\text{(h)}(x)$ is the general solution to the corresponding homogeneous equation, while $\rhoES^\text{(p)}(x)$ is a particular solution.

Looking for solutions of the form $C e^{-x/\lambda}$ for the homogeneous part, we find
\begin{align}
    C \left( \frac{D}{\lambda^2} - \koffS \right) e^{-x/\lambda} = 0. 
\end{align}
The two possible roots for $\lambda$ are $\pm \sqrt{D/\koffS}$. Calling the positive root $\lambdaES$, which represents the mean distance traveled by the substrate--bound enzyme before releasing the substrate, we can write the general solution to the homogeneous part of Eq.~\ref{eqnS:main} as
\begin{align}
    \rhoES^\text{(h)}(x) = C_1 e^{-x/\lambdaES} + C_2 e^{x/\lambdaES},
\end{align}
where $C_1$ and $C_2$ are constants which will be determined from the boundary conditions.

Since the nonhomogeneous part of Eq.~\ref{eqnS:main} is a scaled exponential, we look for a particular solution of the same functional form, namely, $\rhoES^\text{(p)}(x) = C_{\text{p}} e^{-x/\lambdaS}$. Substituting this form into the ODE, we obtain
\begin{align}
    \label{eqn:Cp}
    C_{\text{p}} \left( \frac{D}{\lambdaS^2} - \koffS \right) e^{-x/\lambdaS} = -\kon  \rhoS(0) e^{-x/\lambdaS} \rhoE.
\end{align}
The constant coefficient $C_{\text{p}}$ can then be found as
\begin{align}
    \label{eqn:Cparticular}
    C_{\text{p}} &= \frac{\kon \rhoS(0) \rhoE}{\koffS - \dfrac{D}{\lambdaS^2}} 
    = \frac{\kon \rhoS(0) \rhoE}{\koffS \left( 1 - \dfrac{D/\koffS}{\lambdaS^2} \right)} \nonumber\\
    &= \frac{\kon \rhoS(0) \rhoE}{\koffS \left(1 - \dfrac{\lambdaES^2}{\lambdaS^2} \right)},
\end{align}
where we have 
used the equality $\lambdaES = \sqrt{D/\koffS}$.

Now, to find the unknown coefficients $C_1$ and $C_2$, we impose the no-flux boundary conditions for the density $\rhoES(x)$ at the left and right boundaries of the compartment, namely,
\begin{align}
    \frac{\mathrm{d} \rhoES}{\mathrm{d} x} \big|_{x=0} &= -\frac{C_1}{\lambdaES} + \frac{C_2}{\lambdaES} - \frac{C_{\text{p}}}{\lambdaS} = 0, \\
    \label{eqn:SI_no_flux_r}
    \frac{\mathrm{d} \rhoES}{\mathrm{d} x} \big|_{x=L} &= -\frac{C_1}{\lambdaES} e^{-\frac{L}{\lambdaES}} +  \frac{C_2}{\lambdaES} e^{\frac{L}{\lambdaES}} - \frac{C_{\text{p}}}{\lambdaS} e^{-\frac{L}{\lambdaS}} = 0.
\end{align}
Note that we did not take into account the product formation flux at the rightmost boundary when writing Eq.~\ref{eqn:SI_no_flux_r} in order to simplify our calculations.
This is justified in the limit of slow catalysis -- an assumption that we make in our treatment.
The above system of two equations can then be solved
for $C_1$ and $C_2$, yielding
\begin{align}
    \label{eqn:C1}
    C_1 &= 
    -\frac{\lambdaES}{2\lambdaS} \frac{e^{L/\lambdaES}-e^{-L/\lambdaS}}{\sinh(L/\lambdaES)} C_{\text{p}}, \\
    \label{eqn:C2}
    C_2 &= \frac{\lambdaES}{2\lambdaS} \frac{e^{-L/\lambdaS}-e^{-L/\lambdaES}}{\sinh(L/\lambdaES)} C_{\text{p}}.
\end{align}

With the constant coefficients known, we obtain the general solution for the complex profile as
\begin{align}
    \rhoES(x) &= C_1 e^{-x/\lambdaES} + C_2 e^{x/\lambdaES} + C_{\text{p}} e^{-x/\lambdaS} \nonumber\\
    &= C_{\text{p}} \left( \frac{\lambdaES}{\lambdaS \sinh(L/\lambdaES)} \left[- \frac{e^{(L-x)/\lambdaES} + e^{(x-L)/\lambdaES}}{2} + \frac{e^{-x/\lambdaES} + e^{x/\lambdaES}}{2} e^{-L/\lambdaS} \right] + e^{-x/\lambdaS}\right) \nonumber\\
    &= \frac{\kon \rhoS(0) \rhoE}{\koffS \left(1 - \lambdaES^2/\lambdaS^2 \right)} \left( \frac{\lambdaES}{\lambdaS \sinh(L/\lambdaES)} \left[- \cosh\left(\dfrac{L-x}{\lambdaES}\right) + \cosh\left(\dfrac{x}{\lambdaES}\right) e^{-L/\lambdaS} \right] + e^{-x/\lambdaS}\right) \nonumber\\
    \label{eqn:rhoES_generic}
    &= \frac{\kon \rhoS(0) \rhoE}{\koffS \left(1 - \lambdaES^2/\lambdaS^2 \right)} \left( \frac{\lambdaES}{\lambdaS \sinh(L/\lambdaES)} \left[- \cosh\left(\dfrac{L-x}{\lambdaES}\right) + \cosh\left(\dfrac{x}{\lambdaES}\right) e^{-L/\lambdaS} \right] + e^{-x/\lambdaS}\right).
\end{align}

\subsection{2. Density profile in low and high substrate localization regimes}

If substrate localization is very poor ($\lambdaS \gg L$), the substrate distribution will be uniform ($\rho_S(x) = \bar{\rho}_S =  S_{\text{total}}/L$), resulting in a similarly flat profile of enzyme--substrate complexes with their density $\rhoES^{\infty}$ given by
\begin{align}
    \rhoES^{\infty} &= \frac{\kon \rhoS(0) \rhoE}{\koffS} \nonumber\\
    \label{eqn:rhoES_equil}
    &= \frac{\kon \bar{\rho}_{\text{S}} \rhoE}{\koffS}.
\end{align}
This is the expected equilibrium result where the complex concentration is inversely proportional to the dissociation constant ($\koffS/\kon$).

In the opposite limit where the substrates are highly localized
($\lambdaS \ll \lambdaES, L$ and $\rhoS(0) \approx S_{\text{total}}/\lambdaS$ from Eq.~\ref{eqn:rhoS0}), the complex density profile simplifies into
\begin{align}
    \rhoES(x) &\approx \frac{\kon S_{\text{total}} \rhoE}{\koffS 
    \lambdaS (-\lambdaES^2/\lambdaS^2)} \left( - \frac{\lambdaES}{\lambdaS \sinh(L/\lambdaES)} \cosh\left( \frac{L-x}{\lambdaES} \right) \right) \nonumber\\
    &= \frac{\kon S_{\text{total}} \rhoE}{\koffS L} \frac{L/\lambdaES}{\sinh(L/\lambdaES)} \cosh \left( \frac{L-x}{\lambdaES}\right) \nonumber\\
    \label{eqn:rhoES_ellS_0}
    &= \rhoES^{\infty} \times \frac{L/\lambdaES}{\sinh(L/\lambdaES)} \cosh \left( \frac{L-x}{\lambdaES}\right).
\end{align}
The $x$-dependence through the $\cosh(\cdot)$ function suggests that the complex density is the highest at the leftmost boundary and lowest at the rightmost boundary, with the degree of complex localization dictated by the length scale parameter $\lambdaES$. Notably, this localization of complexes does not alter their total number, since the average complex density is conserved, that is,
\begin{align}
    \langle \rhoES \rangle &= \int_{0}^L \rhoES(x)\, \mathrm{d}x \nonumber\\
    &= \rhoES^{\infty} \times \frac{L/\lambdaES}{\sinh(L/\lambdaES)} \times \frac{1}{L} \int_{0}^L  \cosh \left( \frac{L-x}{\lambdaES} \right) \, \mathrm{d}x \nonumber\\
    &= \rhoES^{\infty} \times \frac{L/\lambdaES}{\sinh(L/\lambdaES)} \times \frac{\lambdaES}{L} \sinh(L/\lambdaES) \nonumber\\
    &= \rhoES^{\infty}.
\end{align}

Eq.~\ref{eqn:rhoES_ellS_0} for the complex profile can be alternatively written in terms of the diffusion time scale $\tau_D = L^2/D$ and the substrate off-rate $\koffS$. Noting that $L/\lambdaES = \sqrt{L^2 \koffS/D} = \sqrt{\tau_D \koffS}$ and introducing a dimensionless coordinate $\tilde{x} = x/L$, we find
\begin{align}
    \label{eqn:rhoES_alt}
    \rhoES(x) = \rhoES^\infty \times \frac{\sqrt{\tau_D \koffS}}{\sinh \left(\sqrt{\tau_D \koffS} \right)} \cosh \left(\sqrt{\tau_D \koffS}(1-\tilde{x}) \right).
\end{align}
The above equation is what was used for generating the plots in Fig.~3b of the main text for different choices of the diffusion time scale.

\subsection{3. Fidelity in low and high substrate localization regimes}

Let us now evaluate the fidelity of the model in the two limiting regimes discussed earlier. In the poor substrate localization case, which corresponds to an equilibrium setting, the fidelity can be found from Eq.~\ref{eqn:rhoES_equil} as
\begin{align}
    \label{eqn:eta_equil}
    \eta_\text{eq} = \frac{r\rhoER^\infty}{r\rhoEW^\infty} = \frac{\koffW}{\koffR},
\end{align}
where we have employed the assumption about the right and wrong substrates having identical density profiles. This is the expected result for equilibrium discrimination where no advantage is taken of the system's spatial structure.

In the regime with high substrate localization, the enzyme--substrate complexes have a nonuniform spatial distribution. What matters for product formation is the complex density at the rightmost boundary ($\tilde{x} = 1$), which we obtain from Eq.~\ref{eqn:rhoES_alt} as
\begin{align}
    \rhoES(L) = \rhoES^\infty \times \frac{\sqrt{\tau_D \koffS}}{\sinh \left(\sqrt{\tau_D \koffS} \right)}.
\end{align}
Substituting the above expression written for right and wrong complexes into the definition of fidelity, we find 
\begin{align}
    \eta &= \frac{r \rhoER(L)}{r\rhoEW(L)} \nonumber\\
    &= \eta_{\text{eq}} \times \sqrt{\frac{\koffR}{\koffW}} \frac{\sinh\left(\sqrt{\tau_D \koffW} \right)}{\sinh\left(\sqrt{\tau_D \koffR} \right)} \nonumber\\
    \label{eqn:eta_ideal_seq}
    &= \sqrt{\eta_{\text{eq}}} \frac{\sinh\left(\sqrt{\tau_D \koffW} \right)}{\sinh\left(\sqrt{\tau_D \koffR} \right)}.
\end{align}

This is the result reported in Eq.~5 of the main text.
To gain more intuition about it and draw parallels with traditional kinetic proofreading, let us consider the limit of long diffusion time scales where proofreading is the most effective. In this limit, the hyperbolic sine functions above can be approximated as $\sinh(\scalebox{0.8}{$\sqrt{\tau_D \koffS}$}) \approx 0.5 \, e^{\sqrt{\tau_D \koffS}}$, simplifying the fidelity expression into
\begin{align}
    \eta &= \sqrt{\eta_{\text{eq}}}  \frac{e^{\sqrt{\tau_D \koffW}}}{e^{\sqrt{\tau_D \koffR}}} \nonumber\\
    &= \sqrt{\eta_{\text{eq}}} e^{\sqrt{\tau_D \koffW} - \sqrt{\tau_D \koffR}} \nonumber\\
    \label{eqn:eta_eta_eq}
    &= \sqrt{\eta_{\text{eq}}} e^{\sqrt{\tau_D \koffR} \left( \sqrt{\eta_\text{eq}} - 1\right)},
\end{align}
where we have used the definition of equilibrium fidelity (Eq.~\ref{eqn:eta_equil}). In traditional proofreading, a scheme with $n$ proofreading realizations can yield a maximum fidelity of $\eta/\eta_{\text{eq}} = \eta_{\text{eq}}^n$. The value of $n$ for the original Hopfield model, for instance, is $1$. It would be informative to also know the effective parameter $n$ for the spatial proofreading model. Dividing Eq.~\ref{eqn:eta_eta_eq} by $\eta_{\text{eq}}$, we find
\begin{align}
    \frac{\eta}{\eta_{\text{eq}}} = \frac{1}{\sqrt{\eta_{\text{eq}}}} e^{\sqrt{\tau_D \koffR} \left( \sqrt{\eta_\text{eq}} - 1\right)} &= \eta_{\text{eq}}^n, \nonumber\\
    e^{\sqrt{\tau_D \koffR} \left( \sqrt{\eta_\text{eq}} - 1\right)} &= \eta_{\text{eq}}^{n+\frac{1}{2}}, \nonumber\\
    \sqrt{\tau_D \koffR} \left( \sqrt{\eta_\text{eq}} - 1\right) &= \left( n+ \frac{1}{2}\right) \ln \eta_{\text{eq}} \Rightarrow \nonumber\\
    n + \frac{1}{2} &= \frac{\sqrt{\eta_{\text{eq}}} - 1}{\ln \eta_{\text{eq}}} \sqrt{\tau_D \koffR}.
\end{align}
This exact result can be simplified into an approximate form when diffusion is slow and $\eta_{\text{eq}} \gg 1$, yielding the expression reported in Eq.~7 of the main text, namely,
\begin{align}
    n &\approx \frac{\sqrt{\eta_{\text{eq}}} \sqrt{\tau_D \koffR}}{\ln \eta_{\text{eq}}} \nonumber\\
    &= \frac{\sqrt{\tau_D \koffW}}{\ln \eta_{\text{eq}}}.
\end{align}

\subsection{4. Fidelity in an intermediate substrate localization regime}

The generic expression for complex density at the rightmost boundary ($x=L$) can be written using Eq.~\ref{eqn:rhoES_generic} as
\begin{align}
    \label{eqn:rhoES_L_general}
    \rhoES(L) = \frac{\kon \rhoS(0) \rhoE}{\koffS \left(1 - \lambdaES^2/\lambdaS^2 \right)} \left( \frac{\lambdaES}{\lambdaS \sinh(L/\lambdaES)} \left[\cosh\left(\dfrac{L}{\lambdaES}\right) e^{-L/\lambdaS} - 1 \right] + e^{-L/\lambdaS}\right).
\end{align}
For the system to proofread, substrates need to be sufficiently localized ($\lambdaS < L$) and diffusion needs to be sufficiently slow ($\tau_D \koffS > 1$ or, $\lambdaES < L$). Under these conditions, the substrate profile can be approximated using Eq.~\ref{eqn:rhoS0} as $\rhoS(x) \approx \lambdaS^{-1} S_{\text{total}}e^{-x/\lambdaS}$, while the hyperbolic sine and cosine functions used above can be approximated as $\sinh(L/\lambdaES) \approx \cosh(L/\lambdaES) \approx 0.5\, e^{L/\lambdaES}$. With these approximations, the complex density expression simplifies into
\begin{align}
    \rhoES(L) &= \frac{\kon S_{\text{total}} \rhoE}{\koffS \lambdaS \left(1 - \lambdaES^2/\lambdaS^2 \right)} \left( \frac{\lambdaES}{\lambdaS} \left[e^{-L/\lambdaS} - 2 e^{-L/\lambdaES} \right] + e^{-L/\lambdaS}\right) \nonumber\\
    &= \frac{\kon S_{\text{total}} \rhoE}{\koffS (\lambdaS^2 - \lambdaES^2)} \left( (\lambdaS + \lambdaES)e^{-L/\lambdaS} - 2\lambdaES e^{-L/\lambdaES}\right).
\end{align}
Now, depending on how $\lambdaS$ compares with $\lambdaES$, there can be two qualitatively different regimes for the complex density, namely,
\begin{align}
    \label{eqn:rhoES_cases}
    \rhoES(L) = \rhoES^\infty \times
    \begin{cases}
        \dfrac{2L}{\lambdaES} e^{-L/\lambdaES}, &\quad \text{if } \lambdaS \ll \lambdaES \, \, \,  \color{gray}{\left(L/\lambdaS \gg \sqrt{\tau_D \koffS} \right)} \\[12pt]
        \dfrac{L}{\lambdaS} e^{-L/\lambdaS}, &\quad \text{if } \lambdaES \ll \lambdaS \, \, \,  \color{gray}{\left(\sqrt{\tau_D \koffS} \gg L/\lambdaS \right)}
    \end{cases}
\end{align}
where we used the equilibrium complex density $\rhoES^\infty$ defined in Eq.~\ref{eqn:rhoES_equil}.

Notably, the first regime effectively corresponds to the case of ideal sequestration where complex density is independent from the precise value of $\lambdaS$. The dimensionless number $\sqrt{\tau_D \koffS}$ sets the scale for the minimum $L/\lambdaS$ value beyond which ideal sequestration can be assumed. Conversely, the second regime corresponds to the case where the distance traveled by a complex before dissociating is so short that the complex profile is dictated by the substrate profile itself. Because of that, the complex density reduction from its equilibrium limit is independent from the precise values of $\tau_D$ and $\koffS$, as long as the condition $\lambdaES \gg \lambdaS$ is met.

\begin{figure}[!ht]
\includegraphics{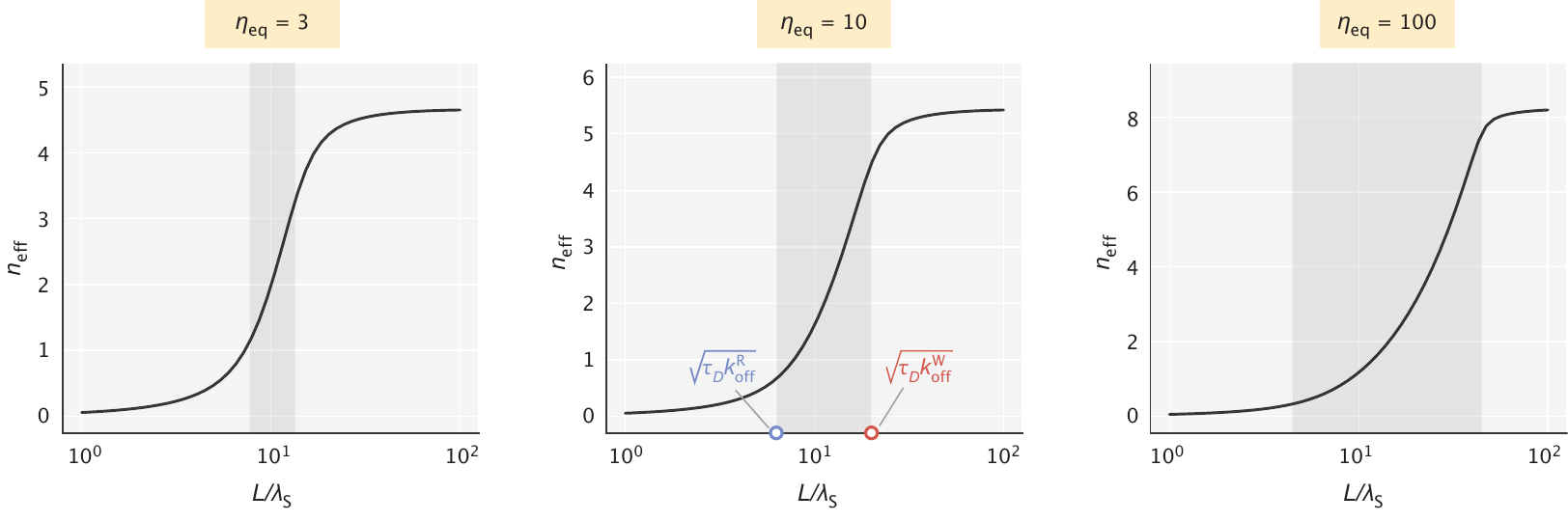}
\caption{The effective number of proofreading realizations ($n_{\text{eff}})$ as a function of $L/\lambdaS$. The shaded region represents the range of $L/\lambdaS$ values set by the key dimensionless numbers $\sqrt{\tau_D \koffR}$ and $\sqrt{\tau_D \koffW}$. $\tau_D$  values chosen for the demonstration were 60, 40, and 20 (in $1/\koffR$ units) for the three different choices of $\eta_{\text{eq}}$, respectively.}
\label{fig:ellS_shadowy}
\end{figure}

The scheme yields its highest fidelity when both right and wrong complex densities are in the first regime (ideal sequestration). When both densities are in the second regime, fidelity is reduced down to its equilibrium value $\eta_{\text{eq}}$ (Table~\ref{table:three_regimes}). The transition between these two extremes happens when the density profiles of right and wrong complexes fall under different regimes. Fidelity can be navigated in the transition zone by tuning the substrate gradient length scale $\lambdaS$. This is demonstrated in Fig.~\ref{fig:ellS_shadowy} for three different choices of $\eta_{\text{eq}}$. In all three cases, the dimensionless numbers $\sqrt{\tau_D \koffR}$ and $\sqrt{\tau_D \koffW}$ set the approximate range in which the bulk of fidelity enhancement occurs, as stated in the main text.

\begin{table}[!ht]
\begin{tabular}{|>{\centering\arraybackslash}m{2cm}|>{\centering\arraybackslash}m{3.2cm}|>{\centering\arraybackslash}m{2cm}|} 
\hline
\rule{0pt}{3.5ex} \rule[-1.7ex]{0pt}{0pt}   & $\lambdaS \ll \lambdaER$ & $\lambdaS \gg \lambdaER$\\
\hline
\hline
\rule{0pt}{5ex} \rule[-3.2ex]{0pt}{0pt} $\lambdaS \ll \lambdaEW$ & $\dfrac{\lambdaEW}{\lambdaER} e^{L\left( \lambdaEW^{-1} - \lambdaER^{-1} \right)}$ & -\\
\hline
\rule{0pt}{5ex} \rule[-3.2ex]{0pt}{0pt} $\lambdaS \gg \lambdaEW$ & $\dfrac{2\lambdaS}{\lambdaER} e^{L \left(\lambdaS^{-1} - \lambdaER^{-1} \right)}$ & $\eta_{\text{eq}}$\\
\hline
\end{tabular}
\caption{Fidelity of the scheme in different regimes of right and wrong complex densities. The upper--right cell is empty because the two conditions on $\lambdaS$ cannot be simultaneously met, since $\lambdaER > \lambdaEW$ by construction (follows from $\koffR < \koffW$).}
\label{table:three_regimes}
\end{table}

\newpage \phantom{blabla}
\newpage
\section*{Appendix B: Energetics of the scheme}

We start this section by deriving an analytical expression for the minimum dissipated power, which was used in making Fig.~4 of the main text. 
Then, we calculate the upper limit on fidelity enhancement available to our model for a finite substrate gradient length scale and compare this limit with the fundamental thermodynamic bound.
We end the section by providing an estimate for the baseline cost of setting up gradients and compare this cost with the maintenance cost reported in the main text.
Similar to our treatment of Appendix A, here too our calculations are based on the 
$\rhoE \approx \text{constant}$ assumption to allow for intuitive analytical results.

\subsection*{1. Derivation of dissipated power}
As stated in the main text, we calculate the minimum rate of energy dissipation necessary for maintaining the substrate profiles as
\begin{align}
    P = \sum_{\text{S=R,W}}\int_{0}^L \jS(x) \mu(x) \mathrm{d}x,
\end{align}
where $\jS(x) = \kon \rhoS(x) \rhoE - \koffS \rhoES(x)$ is the net local substrate binding flux and $\mu(x) = \mu(0) - k_{\text{B}}T \cdot \ln(x/\lambdaS)$ is the local chemical potential. Substituting the analytical expression for $\rhoES(x)$ found earlier (Eq.~\ref{eqn:rhoES_generic}) into $\jS(x)$ and performing a somewhat cumbersome integral, we obtain
\begin{align}
    \beta P = J_{\text{bind}} \sum_{\text{S=R,W}} \frac{1}{1-\lambdaS^2/\lambdaES^2} \left( \frac{\lambdaES}{\lambdaS} \frac{\tanh\left( L/2\lambdaES\right)}{\tanh\left( L/2\lambdaS\right)} -1\right),
\end{align}
where $\beta^{-1} = k_{\text{B}}T$, and $J_{\text{bind}} = \kon \Stot \rhoE$ is the net binding rate of each substrate. Fig.~4 in the main text was made using this expression for power.

To get additional insights about this result, let us consider the case where substrates are highly localized ($\lambdaS \ll L$) and diffusion is slow ($\lambdaES \ll L$) -- conditions needed for effective proofreading. Under these conditions, the hyperbolic tangent terms become 1 and the expression for the power expenditure simplifies into
\begin{align}
    \label{eqn:power_enzyme}
    \beta P = J_{\text{bind}} \sum_{\text{S=R,W}} \frac{\lambdaES^2}{\lambdaS(\lambdaES + \lambdaS)}.
\end{align}
The monotonic increase of power with $\lambdaES$ suggests that energy is primarily spent on maintaining the concentration gradient of right substrates. This is not surprising, since typically right complexes travel a much greater distance into the low concentration region of the compartment before releasing the bound substrate (i.e., $\lambdaER \gg \lambdaEW$). Therefore, neglecting the contribution from wrong substrates and considering the range of $\lambdaS$ values where the bulk of power--fidelity trade-off takes place ($\lambdaER > \lambdaS > \lambdaEW$), we further simplify the power expression into
\begin{align}
    \beta P \approx \frac{J_{\text{bind}} \lambdaER}{\lambdaS} = \frac{J_{\text{bind}} \cdot \beta \Delta \mu}{\sqrt{\tau_D \koffR}},
\end{align}
where we used the identities $\beta \Delta \mu = L/\lambdaS$ and $\lambdaER = L/\sqrt{\tau_D \koffR}$. This simple linear relation suggests that in order to maintain the exponential substrate profile, the minimum energy spent per substrate binding event should be at least $\Delta \mu/\sqrt{\tau_D \koffR} > 1 \, k_{\text{B}}T$.

\subsection*{2. Limits on fidelity enhancement}

The error reduction capacity of the spatial proofreading scheme improves with a greater difference in substrate off-rates, as was demonstrated in Fig.~2 of the main text. At the same time, Fig.~3c showed that the finite length scale of substrate localization (or, finite driving force) sets an upper limit on fidelity enhancement for substrates with fixed off-rates. It is therefore of interest to consider these two features together to find the absolute limit on fidelity enhancement available to our model and then compare it with the fundamental bound set by thermodynamics.

Intuitively, fidelity will be enhanced the most if the density of right complexes does not decay across the compartment, while that of wrong complexes decays maximally. The first condition can be met if diffusion is fast or if the unbinding rate of right substrates is low, in which case we have
\begin{align}
    \label{eqn:rhoER_max}
    \rhoER(L) \approx \rhoER^{\infty},
\end{align}
where $\rhoER^{\infty}$ is the equilibrium density of right complexes. Conversely, when the unbinding rate of wrong substrates is very large, the density of wrong complexes is maximally reduced at the rightmost boundary and can be obtained from Eq.~\ref{eqn:rhoES_L_general} by taking the $\lambdaES \rightarrow 0$ limit, namely,
\begin{align}
    \rhoEW(L) &\approx \frac{\kon \rhoE \rhoS(0) e^{-L/\lambdaS}}{\koffW} \nonumber\\
    &= \frac{\kon \rhoE \Stot e^{-L/\lambdaS}}{\lambdaS \left( 1 - e^{-L/\lambdaS}\right)\koffW}  \nonumber\\
    &= \frac{\kon \rhoE \Stot}{\koffW L} \times \frac{L e^{-L/\lambdaS}}{\lambdaS \left( 1  - e^{-L/\lambdaS}\right)} \nonumber\\
    \label{eqn:rhoEW_min}
    &= \rhoEW^{\infty} \times \frac{\beta \Delta \mu \, e^{-\beta \Delta \mu}}{1 - e^{-\beta \Delta \mu}}.
\end{align}
Here $\rhoEW^{\infty}$ is the equilibrium density of wrong complexes, and $\beta \Delta \mu = L/\lambdaS$ is the effective driving force of the scheme. Taking the ratio of Eqs.~\ref{eqn:rhoER_max} and \ref{eqn:rhoEW_min}, we obtain the largest fidelity enhancement of the scheme for the given driving force, namely,
\begin{align}
    \eta = \frac{\rhoER(L)}{\rhoEW(L)} &= \underbrace{\frac{\rhoER^\infty}{\rhoEW^\infty}}_{\eta_{\text{eq}}} \times \frac{ e^{\beta \Delta \mu} - 1}{\beta \Delta \mu}  \Rightarrow \\
    \label{eqn:eta_enhance_upper}
    \left(\eta / \eta_{\text{eq}} \right)^{\text{max}} &= (e^{\beta \Delta \mu} - 1)/ \beta \Delta \mu.
\end{align}
When $\beta \Delta \mu \gtrsim 1$ (or, $\lambdaS \lesssim L)$, the limit above gets further simplified into
\begin{align}
    \label{eqn:eta_enhance_max_approx}
    \left(\eta / \eta_{\text{eq}} \right)^{\text{max}} \approx e^{\beta \Delta \mu}/ \beta \Delta \mu.
\end{align}

\begin{figure}[!ht]
\includegraphics{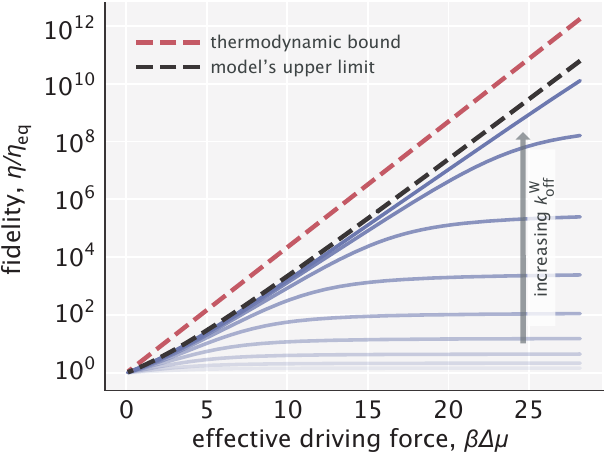}
\caption{Fidelity enhancement as a function of the effective driving force for varying choices of $\koffW$. The red dashed line indicates the thermodynamic bound given by $e^{\beta \Delta \mu}$. The black dashed line corresponds to the model's upper limit on fidelity enhancement given by Eq.~\ref{eqn:eta_enhance_upper}.}
\label{fig:fidelity_limit}
\end{figure}

Now, thermodynamics imposes an upper bound on fidelity enhancement by any proofreading scheme operating with a finite chemical potential $\Delta \mu$. This bound is equal to $e^{\beta \Delta \mu}$ and is reached when the entire chemical potential is used to increase the free energy difference between right and wrong substrates \cite{Qian2006}. Comparing it with the result in Eq.~\ref{eqn:eta_enhance_max_approx}, we can see that fidelity enhancement in the spatial proofreading model has the same exponential scaling term, but with an additional linear factor. Since the dominant contribution comes from the exponential term (as captured also in Fig.~\ref{fig:fidelity_limit}), we can claim that our proposed model can operate very close to the fundamental thermodynamic limit.

\subsection{3. Energetic cost to set up a concentration gradient}

Earlier in the section, we calculated the rate at which energy needs to be dissipated to counteract the homogenizing effect that enzyme activity has on the substrate gradient. In addition to this cost, however, there is also a baseline cost for setting up a gradient in the absence of any enzyme. Here, we calculate this cost in the case where the gradient formation mechanism needs to work against diffusion that tends to flatten the substrate profile.

As before, we consider an exponentially decaying substrate gradient with a decay length scale $\lambdaS$ and a total number of substrates $\Stot$. We write the minimum power $P_D$ required for counteracting the diffusion of substrates as
\begin{align}
    \label{eqn:P_D}
    P_D = - \int_0^L J_D (x) \mu'(x)\, \mathrm{d}x,
\end{align}
where $J_D = -\DS \nabla \rhoS(x)$ is the diffusive flux, with $\DS$ being the substrate diffusion constant. The rationale for writing this form is that diffusion moves substrates from a higher chemical potential region into a neighboring lower chemical potential region. The gradient maintaining mechanism would need to spend at least this chemical potential difference ($\delta \mu = -\mu'(x)\delta x$) per each substrate diffusing a distance $\delta x$ down the chemical potential gradient. Adding up the contribution from all local neighborhoods with a local diffusive flux $J_D(x)$ results in Eq.~\ref{eqn:P_D}.

Now, substituting $\rhoS(x) \sim e^{-x/\lambdaS}$ for the substrate profile and $\mu(x) = \mu(0) + k_{\text{B}}T \ln \left(\text{\scalebox{1.15}{$ \nicefrac{\rhoS(x)}{\rhoS(0)}$}}\right)$ for the chemical potential, we obtain
\begin{align}
    \beta P_D &=  \int_0^L \DS \rhoS'(x) \left( \ln \rhoS(x) \right)' \, \mathrm{d}x \nonumber\\
    &= \DS \int_0^L \frac{\left( \rhoS'(x) \right)^2}{\rhoS(x)} \, \mathrm{d}x \nonumber\\
    &= \DS \int_0^L \frac{\rhoS(x)}{\lambdaS^2} \, \mathrm{d}x \nonumber\\
    \label{eqn:power_P_D}
    &= \frac{\DS \Stot}{\lambdaS^2},
\end{align}
where in the third step we used the relation $\rhoS'(x) = -\rhoS(x)/\lambdaS$. This suggests that the minimum dissipated power required for setting up an exponential gradient increases quadratically with decreasing localization length scale $\lambdaS$.

It is informative to also make a comparison between this result and the earlier calculated minimum dissipation needed to counteract the enzyme's homogenizing activity. Recall that when substrates were sufficiently localized and when diffusion was sufficiently slow, proofreading power could be approximated as (Eq.~\ref{eqn:power_enzyme})
\begin{align}
    \beta P \approx J_{\text{bind}} \frac{\lambdaES^2}{\lambdaS(\lambdaES + \lambdaS)},
\end{align}
where $J_{\text{bind}} = \kon \Stot \rhoE$ is the total substrate binding flux. Using the identities $\lambdaES = \sqrt{D/\koffS}$ and $\KdS = \koffS/\kon$, we can calculate the ratio of the proofreading power to baseline power as
\begin{align}
    \frac{P}{P_D} &= \frac{\kon \Stot \rhoE \lambdaES^2}{\DS \Stot} \times \frac{\lambdaS^2}{\lambdaS \left( \lambdaES + \lambdaS\right)} \nonumber\\
    &= \frac{D}{\DS} \times \frac{\rhoE}{\KdS} \times \frac{\lambdaS/\lambdaES}{1 + \lambdaS/\lambdaES}.
\end{align}
Presuming for simplicity that the enzyme and substrate diffusion constants are the same, we see that two factors determine the power ratio: 1) the amount of free enzyme in the system ($\rhoE/\KdS$), and 2) the substrate localization length scale relative to the characteristic length scale of complex diffusion ($\lambdaS/\lambdaES$). Now, recall that the enzymatic activity on right substrates dominates the proofreading cost (Appendix B1) and that the bulk of fidelity enhancement takes place when $\lambdaS \lesssim \lambdaER$ (Appendix A4). Therefore, when tuning $\lambdaS$ down, initially the power ratio would only depend on the amount of free enzyme in the system ($\rhoE/\KdS$) and then, with tighter substrate localization, the relative contribution of the proofreading power would start to decrease.

In the end, we would like to note that spatial gradients can also be set up using an external potential without a continuous dissipation of energy. In an {\it in vivo} setting, gravity can give rise to spatial structures in oocytes \cite{Feric2013-ga}, while in an {\it in vitro} setting, electric fields can create gradients and power the transport of the complex \cite{Hansen2017--pr}. We leave the investigations of such alternative strategies to future work.

\section*{Appendix C: Studies on the validity of the uniform free enzyme profile  assumption}

\begin{figure}[!ht]
\includegraphics{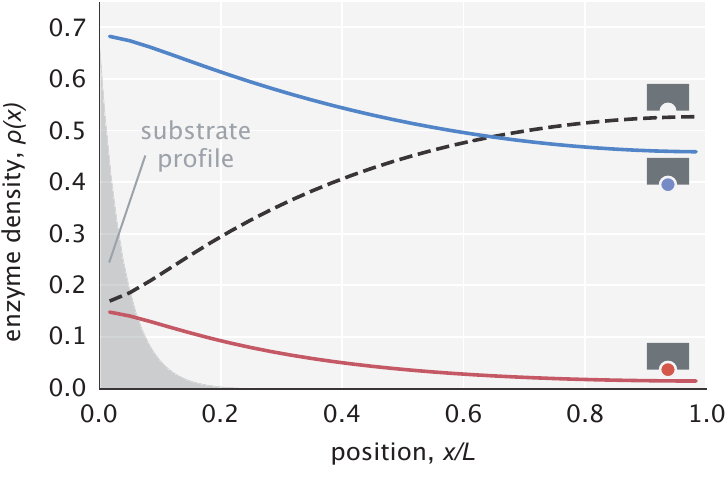}
\caption{Example profiles of free and substrate--bound enzymes. Enzyme profiles are normalized so that the sum of areas under the curves is unity. The substrate profile (rescaled on the $y$-axis) is shown in transparent gray.}
\label{fig:rhoE_variable}
\end{figure}

In our treatment of the model so far, we have assumed for mathematical convenience that free enzymes are in excess, which suggested the approximation $\rhoE(x) \approx \text{constant}$. Example enzyme density profiles shown in Fig.~\ref{fig:rhoE_variable}, however, demonstrate that this assumption does not hold in general. Specifically, there is a depletion of free enzymes near the substrate localization site and abundance near the catalysis site. Because of this depletion at the leftmost edge, we expect a reduction in speed in comparison with our earlier treatment where a flat profile was assumed. In addition, if substrates have a weak gradient, we expect the fidelity to also be reduced, since more enzymes will bind substrates at intermediate positions, reducing the average travel distance to the catalytic site. In what follows, we discuss in greater detail the consequences of having a nonuniform free enzyme distribution on the model performance.

\subsection{1. Effects that relaxing the $\rhoE(x) \approx \text{constant}$ assumption has on the Pareto front}

We begin by studying the effects of relaxing the uniform free enzyme profile assumption on the Pareto front of the speed--fidelity trade-off (Fig.~3a of the main text). This front is reached in the ideal sequestration limit ($\lambdaS \rightarrow 0$). Though in general enzyme profiles need to be obtained using numerical methods due to the nonlinearity of reaction--diffusion equations, in this particular limit ($\lambdaS \rightarrow 0$) an analytical solution is available. To obtain it, we write the reaction--diffusion equations in the bulk region of space as
\begin{align}
    \label{eqn:ER_pde}
    \frac{\partial \rhoER}{\partial t} &= 
    D \frac{\partial^2 \rhoER}{\partial x^2} 
    - \koffR \rhoER\\
    \label{eqn:EW_pde}
    \frac{\partial \rhoEW}{\partial t} &= 
    D \frac{\partial^2 \rhoEW}{\partial x^2} 
    - \koffW \rhoEW\\
    \label{eqn:E_pde}
    \frac{\partial \rhoE}{\partial t} &= 
    D \frac{\partial^2 \rhoE}{\partial x^2}
    + \sum_{\text{S}=\text{R},\text{W}} \koffS \rhoES.
\end{align}
Substrate binding reactions did not enter the above equations, as they occur at the leftmost boundary only. They are instead accounted for via boundary conditions, which read
\begin{align}
    \label{eqn:kon_BC_ER}
    -D \frac{\partial \rhoER}{\partial x}\bigg|_{x=0} &= \kon \Stot \rhoE(0), \\
    \label{eqn:kon_BC_EW}
    -D \frac{\partial \rhoEW}{\partial x}\bigg|_{x=0} &= \kon \Stot \rhoE(0), \\
    \label{eqn:kon_BC_E}
    -D \frac{\partial \rhoE}{\partial x}\bigg|_{x=0} &= -2\kon \Stot \rhoE(0),
\end{align}
where $\Stot$ is the total amount of free substrate of each kind concentrated at $x=0$.

{\underline{Relating local enzyme concentrations.}}
Considering the system at steady state, we add Eqs.~\ref{eqn:ER_pde}-\ref{eqn:E_pde} and obtain
\begin{align}
    0 = D \frac{\mathrm{d}^2 \rhoER}{\mathrm{d} x^2}  + D \frac{\mathrm{d}^2 \rhoEW}{\mathrm{d} x^2}  +  D \frac{\mathrm{d}^2 \rhoE}{\mathrm{d} x^2}.
\end{align}
Diving by $D$ and integrating once, we find
\begin{align}
    \frac{\mathrm{d} \rhoER}{\mathrm{d} x} + \frac{\mathrm{d} \rhoEW}{\mathrm{d} x} +  \frac{\mathrm{d} \rhoE}{\mathrm{d} x} = C_1.
\end{align}
This above relation must hold for arbitrary position $x$. Choosing $x=0$ and noting that from Eqs.~\ref{eqn:kon_BC_ER}-\ref{eqn:kon_BC_E} the sum of fluxes should be zero, we can claim that $C_1 = 0$. Integrating for the second time, we obtain
\begin{align}
    \label{eqn:sum_of_rho}
    \rhoER(x) + \rhoER(x) + \rhoE(x) = C_2,
\end{align}
where $C_2$ is now a different constant. To find it, we perform an integral for the last time across the entire compartment, namely,
\begin{align}
    \int_{0}^L \big(\rhoER(x) + \rhoEW(x) + \rhoE(x) \big) \mathrm{d}x = \Etot = C_2 L.
\end{align}
Here we introduced the parameter $\Etot$ as the total number of enzymes in the system (in free or bound forms). The constant $C_2$, which we will rename into $\rho_0$, is then the average enzyme density, i.e.,
\begin{align}
    \rho_0 = \Etot/L.
\end{align}
Substituting this result into Eq.~\ref{eqn:sum_of_rho}, we find an insightful relation between free and bound enzyme densities at an arbitrary position, namely,
\begin{align}
    \label{eqn:local_profiles}
    \rhoE(x) = \rho_0 - \rhoER(x) - \rhoEW(x).
\end{align}
This relation suggests that whenever the local concentration of bound enzymes is high, the local concentration of free enzymes should be correspondingly low, as we see reflected in the profiles of Fig.~\ref{fig:rhoE_variable}.\\

\underline{Deriving the fidelity expression.} Next, we consider Eqs.~\ref{eqn:ER_pde} and \ref{eqn:EW_pde} separately at steady state, written in the form
\begin{align}
    D \frac{\mathrm{d}^2 \rhoES}{\mathrm{d}x^2} - \koffS \rhoES = 0.
\end{align}
The general solution to this ODE reads
\begin{align}
    \rhoES(x) = C_1^\text{S} e^{-x/\lambdaES} + C_2^\text{S} e^{x/\lambdaES},
\end{align}
where $\lambdaES = \sqrt{D/\koffS}$, and $C_1^\text{S}$ and $C_2^\text{S}$ (S = R,W) are constants which are different for right and wrong complexes. The no-flux boundary condition at $x=L$ can be used to relate these constants and simplify the complex profile expression, namely,
\begin{align}
    -D \frac{\mathrm{d} \rhoES(x)}{\mathrm{d}x} \bigg|_{x=L} &= -\frac{D}{\lambdaES} \left( -C_1^{\text{S}} e^{-L/\lambdaES} + C_2^{\text{S}}e^{L/\lambdaES} \right)= 0 \Rightarrow\\
    C_2^{\text{S}} &= e^{-2L/\lambdaES} C_1^{\text{S}} \Rightarrow \\
    \rhoES(x) &= C_1^{\text{S}}e^{-x/\lambdaES} + C_1^{\text{S}} e^{-2L/\lambdaES}e^{x/\lambdaES} \nonumber\\
    &= 2C_1^{\text{S}} e^{-L/\lambdaES} \cosh\left( \frac{L - x}{\lambdaES} \right) \nonumber\\
    \label{eqn:rhoES_ne_const}
    &= \tilde{C}_1^{\text{S}} \cosh \left( \frac{L-x}{\lambdaES} \right),
\end{align}
where $\tilde{C}_1^{\text{S}} = 2C_1^{\text{S}} e^{-L/\lambdaES}$ is a new constant coefficient introduced for convenience.

Now, the fidelity of the scheme is the ratio of right and wrong complex densities at $x=L$. Using the result above, the fidelity can be written as
\begin{align}
    \label{eqn:eta_ne_ideal}
    \eta = \frac{\rhoER(L)}{\rhoEW(L)} = \frac{\tilde{C}_1^{\text{R}}}{\tilde{C}_1^{\text{W}}}.
\end{align}
The ratio of these constant coefficients can be obtained by noting that the diffusive fluxes of right and wrong complexes at $x=0$ are identical (from Eqs.~\ref{eqn:kon_BC_ER} and \ref{eqn:kon_BC_EW}), that is,
\begin{align}
    -D \frac{\partial \rhoER}{\partial x} \bigg|_{x=0} &= -D \frac{\partial \rhoEW}{\partial x} \bigg|_{x=0} \Rightarrow \\
    \tilde{C}_1^{\text{R}} \times \frac{\sinh(L/\lambdaER)}{\lambdaER} &= \tilde{C}_1^{\text{W}} \times \frac{\sinh(L/\lambdaEW)}{\lambdaEW} \Rightarrow \\
    \frac{\tilde{C}_1^{\text{R}}}{\tilde{C}_1^{\text{W}}} &= \frac{\lambdaER}{\lambdaEW} \frac{\sinh(L/\lambdaEW)}{\sinh(L/\lambdaER)}.
\end{align}
Substituting this result into Eq.~\ref{eqn:eta_ne_ideal}, and recalling the equality $L/\lambdaES = \sqrt{\tau_D \koffS}$, we obtain
\begin{align}
    \eta &= \frac{\sqrt{\tau_D \koffW}}{\sqrt{\tau_D \koffR}} \frac{ \sinh \left(\sqrt{\tau_D \koffW} \right)}{\sinh \left(\sqrt{\tau_D \koffR} \right)} \nonumber\\
    &= \sqrt{\eta_{\text{eq}}} \frac{ \sinh \left( \sqrt{\tau_D \koffW} \right)}{\sinh \left(\sqrt{\tau_D \koffR} \right)}.
\end{align}
This expression is identical to that in Eq.~\ref{eqn:eta_ideal_seq} which was derived under the $\rhoE(x) \approx \text{constant}$ assumption, suggesting that when substrates are highly localized, the shape of the free enzyme profile does not dictate the fidelity.\\

\underline{Deriving the speed expression.} To keep the expression of speed compact while still illustrating the key consequences of relaxing the $\rho(x) \approx \text{constant}$ assumption, we will assume moving forward that the density of wrong complexes is much lower than that of the right complexes, i.e., $\rhoEW(x) \ll \rhoER(x)$. This allows us to approximate the free enzyme density from Eq.~\ref{eqn:local_profiles} as $\rhoE(x) \approx \rho_0 - \rhoER(x)$.

The specification of the right complex density profile requires the knowledge of the unknown coefficient $\tilde{C}_1^{\text{R}}$. To find this coefficient, we use the boundary condition in Eq.~\ref{eqn:kon_BC_ER} and the approximation $\rhoE(x) \approx \rho_0 - \rhoER(x)$ to write
\begin{align}
    D \frac{\tilde{C}^{\text{R}}_1}{\lambdaER} \sinh(L/\lambdaER) &= \kon \Stot \left( \rho_0 - \tilde{C}_1^{\text{R}} \cosh ( L/\lambdaER)\right) \Rightarrow \\
    \tilde{C}_1^{\text{R}} &= \frac{\kon \Stot \rho_0}{\frac{D}{\lambdaER} \sinh(L/\lambdaER) + \kon \Stot \cosh(L/\lambdaER)} \nonumber\\
    &= \frac{\kon \Stot \rho_0}{\lambdaER \koffR \sinh(L/\lambdaER) + \kon \Stot \cosh(L/\lambdaER)} \nonumber\\
    &= \rho_0 \times \frac{\dfrac{\kon \Stot}{\koffR L}}{1 + \dfrac{L}{\lambdaER} \dfrac{\cosh(L/\lambdaER)}{\sinh(L/\lambdaER)} \dfrac{\kon \Stot}{\koffR L}} \times
    \frac{L/\lambdaER}{\sinh(L/\lambdaER)}.
\end{align}
With the constant coefficient known, the right complex density then becomes
\begin{align}
    \label{eqn:rhoER_ne_const}
    \rhoER(x) = \rho_0 \times \frac{\dfrac{\bar{\rho}_{\text{S}}}{\KdR}}{1 + \dfrac{L}{\lambdaER} \dfrac{\cosh(L/\lambdaER)}{\sinh(L/\lambdaER)} \dfrac{ \bar{\rho}_{\text{S}}}{\KdR}} \times
    \frac{L/\lambdaER}{\sinh(L/\lambdaER)} \cosh\left( \frac{L-x}{\lambdaER} \right),
\end{align}
where we used the definitions of the mean substrate density $\bar{\rho}_{\text{S}} = \Stot/L$ and the dissociation constant $\KdR = \koffR/\kon$.

To enable a direct parallel between this general treatment and the earlier one with the $\rhoE(x) \approx \text{constant}$ approximation, let us introduce $\rhoER^\infty$ as the uniform right complex density when diffusion is very fast ($\lambdaER \gg L$) and calculate it from Eq.~\ref{eqn:rhoER_ne_const} as
\begin{align}
    \rhoER^\infty = \rho_0 \times \frac{\dfrac{\bar{\rho}_{\text{S}}}{\KdR}}{1 + \dfrac{\bar{\rho}_{\text{S}}}{\KdR}}.
\end{align}
Now, using the $\rhoER^\infty$ expression, we rewrite Eq.~\ref{eqn:rhoER_ne_const} as
\begin{align}
    \rhoER(x) &= \frac{1 + \dfrac{\bar{\rho}_{\text{S}}}{\KdR}}{1 + \dfrac{L}{\lambdaER} \dfrac{\cosh(L/\lambdaER)}{\sinh(L/\lambdaER)} \dfrac{\bar{\rho}_{\text{S}}}{\KdR}} \times \rhoER^\infty \times  \frac{L/\lambdaER}{\sinh(L/\lambdaER)} \cosh \left( \frac{L-x}{\lambdaER} \right) \nonumber\\
    &= \frac{1 + \dfrac{\bar{\rho}_{\text{S}}}{\KdR}}{1 + \underbrace{\dfrac{L}{\lambdaER} \dfrac{\cosh(L/\lambdaER)}{\sinh(L/\lambdaER)}}_{\gamma} \dfrac{\bar{\rho}_{\text{S}}}{\KdR}} \times \rhoER^{\text{const}}(x),
\end{align}
where $\rhoER^{\text{const}}(x)$ is the complex density obtained under the $\rhoE(x)\approx \text{constant}$ assumption (Eq.~\ref{eqn:rhoES_ellS_0}).
The extra factor that appears on front does not exceed 1 since $\gamma \geq 1$, indicating a reduction in speed, as we anticipated in our more qualitative discussion at the beginning of the section. The presence of the extra factor suggests two possibilities for the approximation to hold true; first, $\gamma \approx 1$ which happens when $\lambdaER \gtrsim L$ or when the right complex does not decay noticeably across the compartment, and second, when $\gamma > 1$ and $\bar{\rho}_{\text{S}} \ll \gamma^{-1} \KdR$, which is when right complexes do decay but their fraction is low compared with free enzymes because of low substrate concentration.\\

\begin{figure}[!ht]
\includegraphics{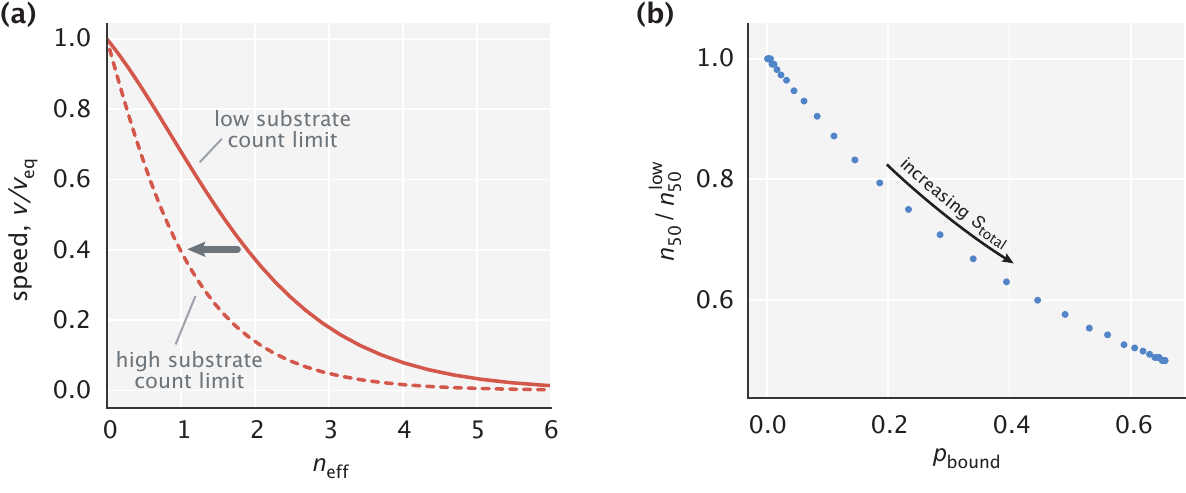}
\caption{Consequences of relaxing the $\rhoE(x) \approx \text{constant}$ assumption on the Pareto front. (a) Pareto fronts in the low and high substrate concentration limits. (b) Reduction in the effective number of proofreading realizations at half--maximum speed as a function of the fraction of enzymes bound. $\eta_{\text{eq}} = 10$ was used in making the plots.}
\label{fig:pareto}
\end{figure}

\underline{Pareto front shift.}
The previous calculations showed that in the ideal substrate sequestration limit relaxing the $\rho(x) \approx \text{constant}$ assumption keeps the fidelity the same while the speed gets reduced. We therefore expect a shift in the Pareto front which is illustrated in Fig.~\ref{fig:pareto}a. To get more intuition about the effect of this shift caused by tuning the amount of substrates, we consider the effective number of proofreading realizations at half--maximum speed ($n_{50}$) and study how this number changes as a function of the fraction of enzymes bound ($p_{\text{bound}} \approx \Etot^{-1} \int \rhoER(x) \, \mathrm{d}x$). Fig.~\ref{fig:pareto}b shows this dependence. As can be seen, $n_{50}$ reduces roughly linearly with $p_{\text{bound}}$; e.g., if 10\% of the enzymes are bound, then a 10\% reduction in $n_{50}$ is expected. This suggests that as long as the fraction of bound enzymes is low, our findings related to the Pareto front made under the $\rhoE\approx \text{constant}$ assumption will generally hold true.

\newpage
\subsection{2. Effects that relaxing the $\rhoE(x) \approx \text{constant}$ assumption has on fidelity in a weak substrate gradient setting}

In this section, we study how accounting for the spatial distribution of free enzymes affects our results on the model's fidelity in the setting where substrates have a finite localization length scale $\lambdaS$. In this setting, Eqs.~(1)-(3) (in the main text) describing the system's dynamics become a system of nonlinear equations, which we solve at steady state using numerical methods.

\begin{figure}[!ht]
\includegraphics{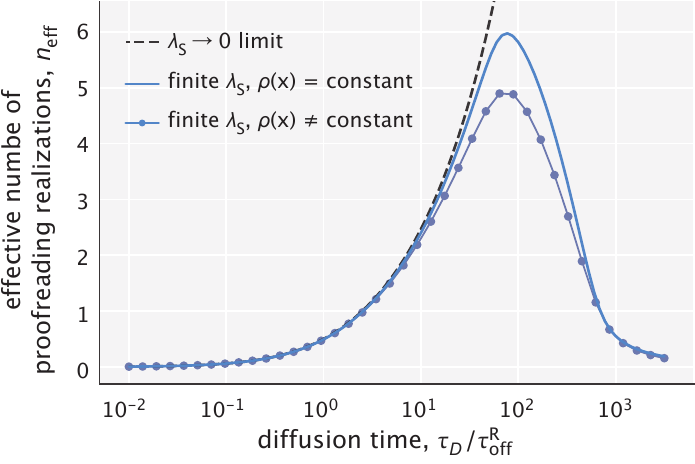}
\caption{
Fidelity as a function of diffusion time scale calculated with and without making the $\rhoE \approx \text{constant}$ approximation. The total number of free substrates is chosen so that 
$\bar{\rho}_{\text{S}}/\KdR = 3$.}
\label{fig:rhoE_variable_tauD}
\end{figure}

An example curve of how fidelity changes with tuning diffusion time scale in a finite $\lambdaS$ setting is shown in Fig.~\ref{fig:rhoE_variable_tauD}. As expected, the nonuniform free enzyme profile leads to a reduction in fidelity. This reduction is not significant when diffusion is relatively fast as in that case the free enzyme profile manages to flatten out rapidly. The reduction is not significant also in the very slow diffusion limit where binding events that lead to production primarily take place in the proximity of the activation region and hence, the nonuniform profile of free enzymes across the compartment has little impact on fidelity. The greatest reduction happens at intermediate diffusion time scales; in particular, when the system achieves its peak fidelity.

\begin{figure}[!ht]
\includegraphics{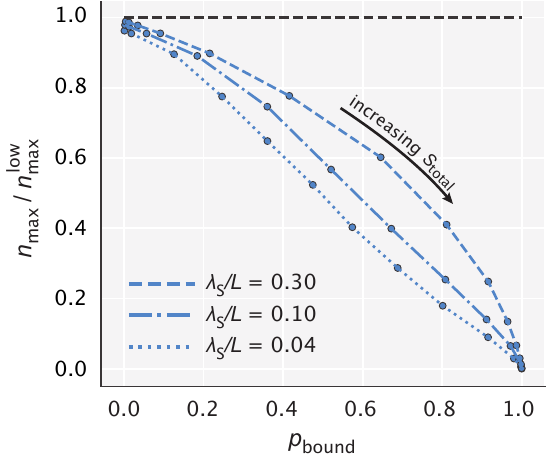}
\caption{Reduction in the peak effective number of proofreading realizations as a function of $p_{\text{bound}}$. $n_{\text{max}}^{\text{low}}$ represents the peak value of $n_{\text{eff}}$ in the limit of low substrate concentration (the maximum of the solid blue curve in Fig.~\ref{fig:rhoE_variable_tauD}).}
\label{fig:rhoE_variable_n_max}
\end{figure}

To quantify the extent of this highest reduction, we calculated the peak value of the effective number of proofreading realizations ($n_{\text{max}}$) for different free substrate amounts which regulate the fraction of bound enzymes ($p_{\text{bound}}$).
The results obtained for different choices of $\lambdaS$ are summarized in Fig.~\ref{fig:rhoE_variable_n_max}.
As can be seen, for the high substrate localization case ($\lambdaS/L = 0.04$), there is a roughly linear dependence between $n_{\text{max}}$ and $p_{\text{bound}}$.
The initial decrease in $n_{\text{max}}$ with growing $p_{\text{bound}}$ is even slower when substrates are less tightly localized ($\lambdaS/L = 0.10, 0.30$).

Taken together, these results suggest that if the substrate concentration is low enough to leave most of the enzymes unbound, then our proposed scheme will proofread efficiently. And this requirement on substrate amount will be further relaxed if diffusion is fast, or if substrates are not very tightly localized.

\newpage
\section*{Appendix D: Proofreading on a kinase/phosphatase-induced gradient}

In this section, we introduce the mathematical modeling setup for the kinase/phosphatase--based gradient formation scheme and describe how its fidelity is calculated numerically. In the end, we discuss the energetics of setting up the substrate concentration gradient and link our calculations to the lower bound on energy cost obtained earlier in Appendix B.

\subsection{1. Setup and estimation of fidelity}

In the analysis thus far, we have imposed a gradient of free substrates and analyzed the proofreading capability of an enzyme acting on this gradient. In a living cell, gradients themselves are maintained by active cellular processes. However, the action of the enzyme  -- that is, binding a substrate in one spatial location, diffusing away, and releasing the substrate elsewhere -- can destroy the gradient, and thereby lead to a loss of proofreading. Here, we analyze the consequences of free substrate depletion and gradient flattening caused by the enzyme.

We model the formation of a substrate gradient by a combination of localized activation and delocalized deactivation. We suppose that substrates can exist in phosphorylated or dephosphorylated forms, and that only the phosphorylated form is capable of binding to the enzyme. The substrates are phosphorylated by a kinase with rate $k_{\text{kin}} = 0.2$ s$^{-1}$, and dephosphorylated by a phosphatase with rate $k_{\text{p}} = 5 $ s$^{-1}$. Crucially, we assume that phosphatases are found everywhere in the domain of size $L \sim 10 \, \text{\textmu m}$ (a typical length scale in a eukaryotic cell), while kinases are localized to one end of the domain (at $x = 0$), as may occur naturally if kinases are bound to one of the membranes enclosing the domain.

The minimal dynamics of phosphorylated substrates and enzyme--substrate complexes is then given by
\bea \label{eq:coupledSubstrateComplexDynamics}
\frac{\partial \rhoS}{\partial t}  \, &&= D \nabla^2 \rhoS - k_{\text{b}} \rhoS + \koffS  \rhoES - k_{\text{p}} \rhoS \nn, \\
\frac{\partial \rhoES}{\partial t}  \, &&= D \nabla^2 \rhoES + k_{\text{b}} \rhoS - \koffS   \rhoES,
\eea
augmented by the boundary conditions
\bea \label{eq:coupledSubstrateComplex_BC}
&& \text{Substrate phosphorylation: } -D \nabla \rhoS \vert_{x = 0} = k_{\text{kin}} \nn, \\
&& \text{No-flux: } -D \nabla \rhoS \vert_{x = L} = -D \nabla \rhoES \vert_{x = L} = -D \nabla \rhoES \vert_{x = 0} = 0.
\eea

Here, we have supposed that the densities of free enzymes, dephosphorylated substrates, and phosphatases are fixed and uniform, and have absorbed them into the relevant rate constants ($k_{\text{b}} = \kon \rhoE$, $k_{\text{kin}}$, and $k_{\text{p}}$, respectively). For simplicity, we have also assumed that the free substrates and enzyme--substrate complexes have the same diffusion coefficient $D = 1 \, \text{\textmu m}^2/\text{s}$.

We numerically solve Eqs. \ref{eq:coupledSubstrateComplexDynamics} and \ref{eq:coupledSubstrateComplex_BC} at steady state. First, the equations of dynamics are made dimensionless by settings units of length and time by $L$ ($\bar{x} = x/L$) and $\tau_D \equiv L^2/D$ ($\bar{t} = t/\tau_D$), respectively. At steady state, the dimensionless equations read
\bea \label{eq:coupledSubstrateComplexDynamics_nondim}
\bar{\nabla}^2 \rhoSbar \, &&= \left(\bar{k}_{\text{b}} + \bar{k}_{\text{p}} \right) \rhoSbar - \koffSbar \rhoESbar \nn, \\
\bar{\nabla}^2 \rhoESbar \, &&=  - \bar{k}_{\text{b}} \rhoSbar + \koffSbar \rhoESbar,
\eea
with boundary conditions
\bea 
\bar{\nabla} \rhoSbar \vert_{\bar{x} = 0} \, &&= - \bar{k}_{\text{kin}}, \nn \\
\bar{\nabla} \rhoSbar \vert_{\bar{x} = 1} \, &&= \bar{\nabla} \rhoESbar \vert_{\bar{x} = 1} = \bar{\nabla} \rhoESbar \vert_{\bar{x} = 0} = 0,
\eea
where concentrations have been rescaled as $\bar{\rho} = \rho L$, and kinetic rates as $\bar{k} = k \, \tau_D$.

We discretize the steady state equations on a grid with spacing $\Delta \bar{x} = 0.01$, approximating the second derivative as
\be \label{eq:secondderiv_FTCS}
\bar{\nabla}^2 \bar{\rho} \approx \f{1}{\Delta \bar{x}^2} \big( \bar{\rho}(\bar{x} + \Delta \bar{x}) + \bar{\rho}(\bar{x} - \Delta \bar{x})  -2 \bar{\rho} (\bar{x}) \big).
\ee

This is ill-defined at the boundaries $\bar{x} = 0$ and $\bar{x} = 1$, which is addressed by incorporating the boundary conditions. For illustration, consider the left boundary, $\bar{x} = 0$, and suppose that our domain included also a point at $\bar{x} = -\Delta \bar{x}$. Then, we could approximate the boundary condition $\bar{\nabla} \rhoSbar \vert_{\bar{x} = 0} = -\bar{k}_{\text{kin}}$ by a centred difference scheme, and solve out for the fictional point at $\bar{x} = -\Delta \bar{x}$, namely,
\bea
&&\bar{\nabla} \rhoSbar \vert_{\bar{x} = 0}  =  - \bar{k}_{\text{kin}} \nn \\
&&\Rightarrow \f{1}{2 \Delta \bar{x}} \big( \rhoSbar(\Delta \bar{x}) - \rhoSbar(-\Delta \bar{x}) \big) = - \bar{k}_{\text{kin}} \nn \\
&&\Rightarrow \rhoSbar(-\Delta \bar{x})  = \rhoSbar(\Delta \bar{x}) + 2 \Delta \bar{x} \, \bar{k}_{\text{kin}}, \nn
\eea
which, when inserted into Eq.~\ref{eq:secondderiv_FTCS}, specifies $\bar{\nabla}^2 \rhoSbar$ at $\bar{x} = 0$, i.e.,
\be
\bar{\nabla}^2 \rhoSbar \vert_{\bar{x} = 0} = \f{1}{\Delta \bar{x}^2} \big( 2\rhoSbar(\Delta \bar{x}) - 2 \rhoSbar(0) \big) + \f{2}{\Delta \bar{x}} \bar{k}_{\text{kin}}.
\ee
Similar considerations apply for the boundary at the right ($\bar{x} = 1$) and for the boundary conditions of $\rhoESbar$.

After discretizing, Eq.~\ref{eq:coupledSubstrateComplexDynamics_nondim} can be written in a matrix form as
\bea
&&\overbrace{\l
\f{1}{\Delta \bar{x}^2}
\begin{pmatrix}
-2 & 2 & 0 & & \cdots & 0 \\
1 & -2 & 1 & & \cdots & 0 \\
\vdots  & \vdots & \vdots & & \ddots & \vdots  \\
0 & \cdots & & 1 & -2 & 1 \\
0 & 0 & \cdots & & 2 & -2 
\end{pmatrix}
- (\bar{k}_{\text{b}} + \bar{k}_{\text{p}}) \mathbf{I} 
\r}^{\mathbf{M}_{\text{S}}}
 \vec{\rho}_{\text{S}} = -\koffSbar  \vec{\rho}_{\text{ES}} +
\overbrace{\begin{pmatrix}
- \f{2}{\Delta \bar{x}} \bar{k}_{\text{kin}} \\ 0 \\\vdots \\0 \\0 
\end{pmatrix}}^{\vec{b}} \nn, \\
&&\underbrace{\l
\f{1}{\Delta \bar{x}^2}
\begin{pmatrix}
-2 & 2 & 0 & & \cdots & 0 \\
1 & -2 & 1 & & \cdots & 0 \\
\vdots  & \vdots & \vdots & & \ddots & \vdots  \\
0 & \cdots & & 1 & -2 & 1 \\
0 & 0 & \cdots & & 2 & -2 
\end{pmatrix}
- \koffSbar  \mathbf{I} 
\r}_{\mathbf{M}_{\text{ES}}}
 \vec{\rho}_{\text{ES}} = - \bar{k}_{\text{b}} \vec{\rho}_{\text{S}},
\eea
where $\vec{\rho}_{\text{S}}$, $\vec{\rho}_{\text{ES}}$ are column vectors of the nondimensionalized concentration profiles evaluated at the spatial grid points, i.e., $\left[ \bar{\rho}(0), \bar{\rho}(\Delta \bar{x}), \cdots \right]^T$. Solving these matrix equations yields
\bea \label{eq:si_biolnumerical}
\vec{\rho}_{\text{S}} \, &&= \l \mathbf{M}_{\text{S}} - \koffSbar \bar{k}_{\text{b}} \mathbf{M}_{\text{ES}}^{-1} \r^{-1} \vec{b}, \nn \\
\vec{\rho}_{\text{ES}} \, &&= - \bar{k}_{\text{b}} \l \mathbf{M}_{\text{S}}\mathbf{M}_{\text{ES}} - \koffSbar \bar{k}_{\text{b}} \mathbf{I} \r^{-1} \vec{b}.
\eea

We compute Eqs.~\ref{eq:si_biolnumerical} numerically for two substrates: a cognate (`R') and a non-cognate (`W'), which differ in their off-rates ($\koffR =0.1\, \text{s}^{-1}$ and $\koffW =1\, \text{s}^{-1}$, respectively). Having the density profiles, the fidelity of the model becomes $\eta \approx \rhoERbar(\bar{x}=1)/\rhoEWbar(\bar{x}=1)$. We calculate the fidelity for different choices of the first--order rate of enzyme--substrate binding ($k_{\text{b}} = \kon \rhoE$); this may be thought of as varying the concentration of free enzyme in the cell. The results are shown in Fig.~5 of the main text.

\subsection{2. Energy dissipation}

In Appendix B3, we estimated the minimum power that a gradient maintaining mechanism would need to dissipate in order to set up an exponentially decaying profile of diffusing substrates. Here, we calculate this power for the kinase/phosphatase--based mechanism and compare it with the lower bound estimated earlier.

Let us assume that phosphorylation and dephosphorylation reactions by kinases and phosphatases are nearly irreversible with associated free energy costs of $\Delta \varepsilon_{\text{kin}}$ and $\Delta \varepsilon_{\text{phosph}}$ per reaction, respectively. The net rate at which active substrates get dephosphorylated is $k_{\text{p}} \Stot$ and it needs to be identical to the net phosphorylation rate of inactive substrates in order for $\Stot$ to remain constant. With the costs of each reaction known, we can write the rate of energy dissipation $P_{\text{k/p}}$ as
\begin{align}
    \label{eqn:P_kp}
    P_{\text{k/p}} = k_{\text{p}} \Stot (\Delta \varepsilon_{\text{kin}} + \Delta \varepsilon_{\text{phosph}}).
\end{align}
Now, when the enzyme activity is very low, the kinase/phosphatase mechanism will create an exponential profile of active substrates with a decay length scale $\lambdaS = \sqrt{\DS/k_{\text{p}}}$. Expressing the rate of phosphorylation in terms of $\lambdaS$ and $\DS$ (i.e., $k_{\text{p}} = \DS/\lambdaS^2$), and substituting it into Eq.~\ref{eqn:P_kp}, we obtain
\begin{align}
    P_{\text{k/p}} = \frac{\DS \Stot}{\lambdaS^2} (\Delta \varepsilon_{\text{kin}} + \Delta \varepsilon_{\text{phosph}}).
\end{align}
Comparing this result with the lower bound found earlier (Eq.~\ref{eqn:power_P_D}), we can note the presence of an extra factor $(\Delta \varepsilon_{\text{kin}} + \Delta \varepsilon_{\text{phosph}})$. Since the free energy consumption during ATP hydrolysis is $\sim10\, k_\text{B}T$, we can say that the dissipated power of the kinase/phosphatase system for setting up an exponential gradient surpasses the lower limit roughly by an order of magnitude.

\bibliography{papers.bib}